\documentclass{sig-alternate-05-2015}
\pdfoutput=1

\usepackage{cite}
\usepackage{algpseudocode}

\usepackage{graphicx}
\usepackage[hang]{subfigure}

\usepackage{caption}
\usepackage{amssymb}
\usepackage{algorithm}
\usepackage{subfig}
\usepackage{blindtext}
\usepackage{yfonts}
\usepackage{comment}
\usepackage{url}
\allowdisplaybreaks
\DeclareMathAlphabet{\mathpzc}{OT1}{pzc}{m}{it}
\usepackage{pifont}
\newcommand*{\inlineequation}[2][]{%
	\begingroup
	\refstepcounter{equation}%
	\ifx\\#1\\%
	\else
	\label{#1}%
	\fi
	\relpenalty=10000 %
	\binoppenalty=10000 %
	\ensuremath{%
		#2%
	}%
	~\@eqnnum
	\endgroup
}

\newtheorem{theorem}{Theorem}

\newtheorem{lemma}{Lemma}
\newtheorem{observation}[theorem]{Observation}

\algtext*{EndFunction}
\algtext*{EndIf}
\algtext*{EndFor}
\algtext*{EndWhile}
\algtext*{EndIf}

\newcommand*{\NINEPAGES}{}

\begin{document}
\newcommand{\eps}{\epsilon}
\newcommand{\set}[1]{\left\{#1\right\}}
\newcommand{\ceil}[1]{ \left\lceil{#1}\right\rceil}
\newcommand{\floor}[1]{ \left\lfloor{#1}\right\rfloor}
\renewcommand{\angle}[1]{ \left\langle{#1}\right\rangle}
\newcommand{\logp}[1]{\log\parentheses{#1}}
\newcommand{\clog}[1]{ \ceil{\log{#1}}}
\newcommand{\clogp}[1]{ \ceil{\logp{#1}} }
\newcommand{\flog}[1]{ \floor{\log{#1}}}
\newcommand{\parentheses}[1]{ \left({#1}\right)}
\newcommand{\abs}[1]{ \left|{#1}\right|}

\newcommand{\cdotpa}[1]{\cdot\parentheses{#1}}
\newcommand{\inc}[1]{$#1 \gets #1 + 1$}
\newcommand{\dec}[1]{$#1 \gets #1 - 1$}
\newcommand{\range}[2][0]{#1,1,\ldots,#2}
\newcommand{\frange}[1]{\set{\range{#1}}}
\newcommand{\xrange}[1]{\frange{#1-1}}
\newcommand{\oneOverE}{ \frac{1}{\eps} }
\newcommand{\oneOverG}{ \frac{1}{\gamma} }
\newcommand{\oneOverT}{ \frac{1}{\tau} }
\newcommand{\smallMultError}{(1+o(1))}
\newcommand{\lowerbound}{\max \set{\log W ,\frac{1}{2\epsilon+W^{-1}}}}
\newcommand{\smallEpsLowerbound}{\window\logp{\frac{1}{\weps}}}
\newcommand{\smallEpsMemoryTheta}{$\Theta\parentheses{\smallEpsMemoryConsumption}$}
\newcommand{\smallEpsMemoryConsumption}{W\cdot\logp{\frac{1}{\weps}}}

\newcommand{\largeEpsRestriction}{For any \largeEps{},}
\newcommand{\largeEps}{$\eps^{-1} \le 2W\left(1-\frac{1}{\logw}\right)$}
\newcommand{\smallEpsRestriction}{For any \smallEps{},}
\newcommand{\smallEps}{$\eps^{-1}>2W\left(1-\frac{1}{\logw}\right)=2\window(1-o(1))$}
\newcommand{\bc}{{\sc Basic-Counting}}
\newcommand{\bs}{{\sc Basic-Summing}}
\newcommand{\windowcounting}{ {\sc $(W,\epsilon)$-Window-Counting}}

\newcommand{\window}{W}
\newcommand{\logw}{\log \window}
\newcommand{\flogw}{\floor{\log \window}}
\newcommand{\weps}{\window\epsilon}
\newcommand{\wt}{\window\tau}
\newcommand{\logweps}{\logp{\weps}}
\newcommand{\logwt}{\logp{\wt}}
\newcommand{\bitarray}{b}
\newcommand{\currentBlockIndex}{i}
\newcommand{\currentBlock}{\bitarray_{\currentBlockIndex}}
\newcommand{\remainder}{y}
\newcommand{\numBlocks}{k}
\newcommand{\sumOfBits}{B}
\newcommand{\blockSize}{\frac{\window}{\numBlocks}}
\newcommand{\iblockSize}{\frac{\numBlocks}{\window}}
\newcommand{\threshold}{\blockSize}
\newcommand{\halfBlock}{\frac{\window}{2\numBlocks}}
\newcommand{\blockOffset}{m}
\newcommand{\inputVariable}{x}

\newcommand{\bcTableColumnWidth}{1.5cm}
\newcommand{\bsTableColumnWidth}{1.7cm}
\newcommand{\bsExtendedTableColumnWidth}{3cm}
\newcommand{\bcExtendedTableColumnWidth}{2.8cm}
\newcommand{\bcNarrowTableColumnWidth}{1.5cm}
\newcommand{\bsNarrowTableColumnWidth}{1.5cm}
\newcommand{\bsWorstCaseTableColumnWidth}{2cm}

\newcommand{\bsrange}{ R }
\newcommand{\bsReminderPercisionParameter}{ \gamma }
\newcommand{\bsest}{ \widehat{\bssum}}
\newcommand{\bssum}{ S^W }
\newcommand{\bsFracInput}{ \inputVariable' }
\newcommand{\bserror}{ \bsrange\window\epsilon }
\newcommand{\bsfractionbits}{ \frac{\bsReminderPercisionParameter}{\epsilon} }
\newcommand{\bsReminderFractionBits}{ \upsilon}
\newcommand{\bsAnalysisTarget}{ \bssum}
\newcommand{\bsAnalysisEstimator}{ \widehat{\bsAnalysisTarget}}
\newcommand{\bsAnalysisError}{ \bsAnalysisEstimator - \bsAnalysisTarget}
\newcommand{\bsRoundingError}{ \xi}


\newcommand{\neps}{\ensuremath{\winSize\eps}}
\newcommand{\Neps}{\ensuremath{\maxWinSize\eps}}
\newcommand{\logn}{\ensuremath{\log\winSize}}
\newcommand{\logN}{\ensuremath{\log\maxWinSize}}
\newcommand{\logneps}{\ensuremath{\logp\neps}}
\newcommand{\logNeps}{\ensuremath{\logp\Neps}}
\newcommand{\oneOverEps}{\ensuremath{\frac{1}{\eps}}}
\newcommand{\winSize}{\ensuremath{n}}
\newcommand{\maxWinSize}{\ensuremath{N}}
\newcommand{\curTime}{\ensuremath{t}}
\newcommand\Tau{\mathrm{T}}
\newcommand{\offset}{\ensuremath{\mathit{offset}}}
\newcommand{\roundedOOE}{k}
\newcommand{\numLargeBlocks}{\frac{\roundedOOE}{4}}
\newcommand{\numSmallBlocks}{\frac{\roundedOOE}{2}}

\newcommand{\remove}{{\sc Remove()}}
\newcommand{\merge}[1]{{\sc Merge(#1)}}
\newcommand{\counting}{{\sc Counting}}
\newcommand{\summing}{{\sc Summing}}
\newcommand{\freq}{{\sc Frequency Estimation}}

\newcommand{\RSS}{Frequent items Algorithm with a Semi-structured Table}
\newcommand{\rss}{FAST}
\newcommand{\slack}{1+\gamma}
\newcommand{\nrCounters}{\ceil{\frac{\slack}{\epsilon}}}
\newcommand{\nrCountersLetter}{\mathfrak{C}}
\newcommand{\step}{\ensuremath{\floor{\frac{M\cdot\gamma}{2} + 1}}}
\newcommand{\stepLetter}{\mathpzc{s}}
\newcommand{\frqEst}{$(\epsilon,M)$-{\sc Volume Estimation}}
\newcommand{\heavyHitters}{$(\theta,\epsilon,M)$-{\sc Weighted Heavy Hitters}}
\newcommand{\query}{{\sc Query$(x)$}}
\newcommand{\winQuery}{{\sc WinQuery$(x)$}}

\newcommand{\SSS}{CSS}
\newcommand{\SpaceS}{Space-Saving}
\newcommand{\CSS}{Compact \SpaceS{}}
\newcommand{\WCSS}{Window \CSS{}}
\newcommand{\SSSInstance}{y}
\newcommand{\queueOfOverflows}{b}
\newcommand{\sumOfOverflows}{B}
\newcommand{\IDArray}{O}
\newcommand{\frameOffset}{o}
\newcommand{\overflowIndicator}{u}
\newcommand{\xFrequency}{v_x}
\newcommand{\xFrequencyEstimator}{\widehat{v_x}}
\newcommand{\xWindowFrequency}{v^W_x}
\newcommand{\xWindowFrequencyEstimator}{\widehat{\xWindowFrequency}}

\newcommand{\streamcounting}{$\epsilon$\textsc{-Counting}}
\newcommand{\paremeterizedStreamcounting}[1]{#1\textsc{-Counting}}
\newcommand{\probabilisticStreamcounting}{$(\epsilon, \delta)$\textsc{-Counting}}
\newcommand{\probabilisticWindowcounting}{$(\window,\epsilon, \delta)$\textsc{-Window Counting}}
\newcommand{\reverseMapping}{$\mathbf R$}

\renewcommand{\gamma}{\phi}
\newcommand{\matrixCellWidth}{5.7cm}
\newcommand{\wideCellWidth}{5.7cm}
\newcommand{\wideCellHeight}{5cm}


\conferenceinfo{ICDCN '18}{January 2018}

\title{Fast Flow Volume Estimation}


\numberofauthors{3}

\author{
\alignauthor
Ran Ben Basat\\
\affaddr{Technion}
\email{sran@cs.technion.ac.il}
\alignauthor
Gil Einziger\\
\affaddr{Nokia Bell Labs}
\email{gil.einziger@nokia.com}
\alignauthor
Roy Friedman\\
\affaddr{Technion}
\email{roy@cs.technion.ac.il}
}


\date{}

\maketitle

\begin{abstract}
The increasing popularity of jumbo frames means growing variance in the size of packets transmitted in modern networks.
Consequently, network monitoring tools must maintain explicit traffic volume statistics rather than settle for packet counting as before.
We present constant time algorithms for volume estimations in streams and sliding windows, which are faster than previous work.
Our solutions are formally analyzed and are extensively evaluated over multiple real-world packet traces as well as synthetic ones.
For streams, we demonstrate a run-time improvement of up to 2.4X compared to the state of the art.
On sliding windows, we exhibit a memory reduction of over 100X on all traces and an asymptotic runtime improvement to a constant.
Finally, we apply our approach to hierarchical heavy hitters and achieve an empirical 2.4-7X speedup.
\end{abstract}


\section{Introduction}
\label{sec:intro}

Traffic measurement is vital for many network algorithms
such as
routing, load balancing, quality of service, caching and anomaly/intrusion
\ifdefined\EXTENDED
detection~\cite{ApproximateFairness,IntrusionDetection,TrafficEngeneering,IntrusionDetection2,LoadBalancing,TinyLFU}.
\else
detection~\cite{ApproximateFairness,IntrusionDetection,IntrusionDetection2,TinyLFU}.
\fi
Typically, networking devices handle millions of flows~\cite{CounterArray1,CounterArray2}.
Often, monitoring applications track the most frequently appearing flows, known as \emph{heavy hitters}, as their impact is most significant.

Most works on heavy hitters identification have focused on packet counting~\cite{WCSS,ICEBuckets,CEDAR}.
However, in recent years jumbo frames and large TCP packets are becoming increasingly popular and so the variability in packet sizes grows.
Consequently, plain packet counting may no longer serve as a good approximation for bandwidth utilization.
For example, in data collected by~\cite{CAIDASJ14} in 2014, less than $1\%$ of the packets account for over $25\%$ of the total traffic.
Here, packet count based heavy hitters algorithms might fail to identify some heavy hitter flows in terms of bandwidth consumption.

Hence, in this paper we explicitly address monitoring of flow \emph{volume} rather than plain packet counting.
Further, given the rapid line rates and the high volume of accumulating data, an aging mechanism such as a \emph{sliding window} is essential for ensuring data freshness and the
\ifdefined \EXTENDED
volume
\fi
estimation's relevance.
Hence, we study estimations of flow volumes in both streams and sliding windows.

Finally, per flow measurements are not enough for certain functionalities like anomaly detection and Distributed Denial of Service (DDoS) attack detection~\cite{HeavyHitters,Sekar2006}.
In such attacks, each attacking device only generates a small portion of the traffic and is not a heavy hitter.
Yet, their combined traffic volume is overwhelming.
\emph{Hierarchical heavy hitters} (HHH) aggregates traffic from IP addresses that share some common prefix~\cite{RHHH}.
In a DDoS, when attacking devices share common IP prefixes, HHH can discover the attack.
To that end, we consider volume based HHH detection as well.



Before explaining our contribution, let us first motivate why packet counting solutions are not easily adaptable to volume estimation.
Counter algorithms typically maintain a fixed set of counters~\cite{SpaceSavings,RAP,LC,PLC,MLC,BatchDecrement,WCSS} that is considerably smaller than the number of flows.
Ideally, counters are allocated to the heavy hitters. 
When a packet from an unmonitored flow arrives, the corresponding flow is allocated the minimal counter~\cite{SpaceSavings} or a counter whose value has dropped below a dynamically increased threshold~\cite{LC}.

We refer to a stream in which each packet is associated with a\emph{weight} is as a \emph{weighted} stream. Similarly, we refer to streams without weights, or when all packets receive the same weight as \emph{unweighted}. 
For unweighted streams, ordered data structures allow constant time updates and queries~\cite{SpaceSavings,WCSS}, since when a counter is incremented, its relative order among all counters changes by at most one.
Unfortunately, maintaining the counters sorted after a counter increment in a weighted stream either requires to search for its new location, which incurs a logarithmic cost, or resorting to logarithmic time data structures like heaps.
The reason is that if the counter is incremented by some value $w$, its relative position might change by up to $w$ positions.
This difficulty motivates our work\footnote{
The most naive approach treats a packet of size $w$ as $w$ consecutive arrivals of the same packet in the unweighted case, resulting in linear update times, which is even worse.}.


\subsection{Contributions}
We contribute to the following network
\ifdefined\EXTENDED
traffic
\fi
measurement problems: (i) stream heavy hitters, (ii) sliding window heavy hitters, (iii) stream hierarchical heavy hitters.
Specifically, our first contribution is \emph{\RSS} (\rss), a novel algorithm for monitoring flow volumes and finding heavy hitters.
\rss{} processes elements in worst case $O(1)$ time using asymptotically optimal space.
\ifdefined\EXTENDED
A major part of our contribution lies in the detailed formal analysis we perform, which proves the above properties, as well as in and the accompanying performance study.
We evaluate \rss{} on five real Internet packet traces from a data center and backbone networks.
We demonstrate a speedup of up to a factor of 2.4X compared to previous works.
\fi
\ifdefined\NINEPAGES
We formally prove and analyze the performance of \rss.
We then evaluate \rss{} on 5 real Internet packet traces from a data center and backbone networks, demonstrating a 2.4X performance gain compared to previous works.
\fi

Our second contribution is \emph{Windowed \RSS} (W\rss), a novel algorithm for monitoring flow volumes and finding heavy hitters in sliding windows.
We evaluate W\rss{} on five Internet traces and show that its runtime is reasonably fast, and that it requires as little as $1\%$ of the memory of previous work~\cite{HungAndTing}.
We analyze W\rss{} and show that it operates in constant time and is space optimal, which asymptotically improves both the runtime and the space consumption of previous work.
We believe that such a dramatic improvement makes volume estimation over a sliding window practical!

Our third contribution is \emph{Hierarchical \RSS} (H\rss), which finds hierarchical heavy hitters.
H\rss{} is created by replacing the underlying HH algorithm in~\cite{HHHMitzenmacher} (Space Saving) with \rss.
We evaluate H\rss{} and demonstrate an asymptotic update time improvement as well as an empirical 2.4-7X speedup on real Internet traces.

\section{Related Work}
\label{sec:related}
\ifdefined\EXTENDED
As mentioned above, our work addresses three related problems, which we survey below.
\fi
\subsection{Streams}
\label{subsec:streams}

\ifdefined\EXTENDED
\subsubsection{Estimators and Sampling}
\emph{Probabilistic short counters}, or \emph{estimators}, represent large numbers using small counters by degrading 
precision~\cite{CEDAR,ICEBuckets,CASE}.
By shrinking counters' size, more flows can be monitored in SRAM.
But these methods still require maintaining a flow-to-counter mapping that often requires more space than the counters themselves.
Sampling is also an attractive approach when space is scarce~\cite{Sample1,Sample2,Sample3} despite the resulting sampling error.
\fi

Sketches such as \emph{Count Sketch (CS)}~\cite{CountSketch} and \emph{Count Min Sketch (CMS)}~\cite{CMSketch} are attractive as they enable counter sharing and need not maintain a flow to counter mapping for all flows.
Sketches typically only provide a probabilistic estimation, and often do not store flow identifiers. Thus, they cannot find the heavy hitters, but only focus on the volume estimation problem.
Advanced sketches, such as Counter Braids~\cite{CounterBraids}, Randomized Counter Sharing~\cite{RandomizedCounterSharing} and Counter Tree~\cite{CounterTree}, improve 
\ifdefined\EXTENDED
accuracy but their queries require complicated decoding procedures that can only be done off-line.
\else
accuracy, but their queries require complex decoding.
\fi

In \emph{counter based} algorithms, a flow table is maintained, but only a small number of flows are monitored.
These algorithms differ from each other in the size and maintenance policy of the flow table, e.g.,~\emph{Lossy Counting}~\cite{LC} and its extensions~\cite{PLC,MLC}, \emph{Frequent}~\cite{BatchDecrement} and \emph{Space Saving}~\cite{SpaceSavings}.
Given ideal conditions, counter algorithms are considered superior to sketch based techniques.
Particularly, Space Saving was empirically shown to be the most accurate~\cite{SpaceSavingIsTheBest,SpaceSavingIsTheBest2010,SpaceSavingIsTheBest2011}.
Many counter based algorithms were developed by the databases community and are mostly suitable for software implementation.
The work of~\cite{WCSS} suggests a compact static memory implementation of Space Saving that may be more accessible for hardware design.
Yet, software implementations are becoming increasingly relevant in networking as emerging technologies such as NFVs become popular.

Alas, most previous works rely on sorted data structures such as \emph{Stream Summary}~\cite{SpaceSavings} or SAIL~\cite{WCSS} that only operate in constant time for unweighted updates. 
\ifdefined\EXTENDED
As mentioned, existing sorted data structures cannot be maintained in constant time in the weighted updates case.
\fi
Thus, a logarithmic time heap based implementation of Space Saving was suggested~\cite{SpaceSavingIsTheBest2010} for the more general volume counting problem.
IM-SUM, DIM-SUM~\cite{IM-SUM} and BUS-SS~\cite{BUS} are very recent algorithms developed for the volume heavy-hitters problem (\emph{only} for streams, with \emph{no} sliding windows support). BUS offers a randomized algorithm that operates in constant time. IM-SUM operates in amortized $O(1)$ time and DIM-SUM in worst case constant time. Empirically, DIM-SUM it is slower than FAST.
Additionally, DIM-SUM requires $\frac{2+\gamma}{\epsilon}$ counters, for some $\gamma>0$, for guaranteeing $N\cdot M\cdot \epsilon$ error and operating in $O(\gamma^{-1})$ time. FAST only needs half as many counters for the same time and error guarantees. 

\subsection{Sliding Windows}
\label{subsec:slidingWindows}

Heavy hitters on sliding windows were first studied by~\cite{ArasuM04}. 
Given an accuracy parameter $(\varepsilon)$, a window size $(W)$ and a maximal increment size ($M$), such algorithms estimate flows' volume on the sliding window with an additive error that is at most $W\cdot M\cdot \varepsilon$.

Their algorithm requires $O\parentheses{\frac{1}{\eps}\log^2\frac{1}{\eps}}$ counters and $O\parentheses{\frac{1}{\eps}\log\frac{1}{\eps}}$ time for queries and updates.
The work of~\cite{LeeT06} reduces the space requirements and update time to $O\parentheses{\frac{1}{\eps}}$.
An improved algorithm with a constant \emph{update time} is given in~\cite{HungLT10}.
Further, \cite{WCSS} provided an algorithm that requires $O\parentheses{\frac{1}{\eps}}$ for queries and supports constant time updates and item frequency queries.

The weighted variant of the problem was only studied by~\cite{HungAndTing}, whose algorithm operates in $O\parentheses{\frac{A}{\epsilon}}$ time and requires $O\parentheses{\frac{A}{\epsilon}} $ space for a $W\cdot M\cdot \varepsilon$ approximation; here, $A\in[1,M]$ is the \emph{average} packet size in the window.
In this work, we suggest an algorithm for the weighted problem that ($i$) uses optimal $O\parentheses{\frac{1}{\eps}}$ space, ($ii$) performs heavy hitters queries in optimal $O\parentheses{\frac{1}{\eps}}$ time, and (iii)
performs volume queries and updates in constant time.

\subsection{Hierarchical Heavy Hitters}
\label{subsec:HHH}
\emph{Hierarchical Heavy Hitters (HHH)} were 
\ifdefined\EXTENDED
first defined by~\cite{Cormode2003}, and then extended to multiple dimensions in~\cite{Cormode2004,CormodeHHH,Hershberger2005,HeavyHitters,HHHMitzenmacher}.
\else
addressed, e.g., in~\cite{Cormode2004,CormodeHHH,Hershberger2005,HeavyHitters,HHHMitzenmacher}.
\fi
HHH algorithms monitor aggregates of flows that share a common prefix.
To do so, HHH algorithms treat flows identifiers as a hierarchical domain. We denote by $H$ the size of this domain.

\ifdefined\EXTENDED
A single dimension algorithm requiring $O\left(\frac{H^2}{\epsilon}\right)$ space 
was introduced in~\cite{Lin2007}.
Later,~\cite{Truong2009} showed a two dimensions algorithm requiring $O\left(\frac{H^{3/2}}{\epsilon}\right)$ space and update time.
\fi
The full and partial ancestry algorithms~\cite{CormodeHHH} are trie based algorithms that require $O\left(\frac{H}{\epsilon} \log \epsilon N \right)$ space and operate at $O\left(H \log{\epsilon N}  \right)$ time.
The state of the art~\cite{HHHMitzenmacher} algorithm requires $O\left(\frac{H}{\epsilon}\right)$ space and its update time for weighted inputs is $O\left(H \log(\frac{1}{\epsilon})\right)$.
\ifdefined\EXTENDED
The algorithm of~\cite{HHHMitzenmacher}
\else
It
\fi
solves the approximate HHH problem by dividing it into multiple simpler heavy hitters problems.
In our work, we replace the underlying heavy hitters algorithm of~\cite{HHHMitzenmacher} with \rss, which yields a space complexity of $O\left(\frac{H}{\epsilon}\right)$ and an update complexity of $O(H)$.
That is, we improve the update complexity from $O\left(H \log\left(\frac{1}{\epsilon}\right)\right)$ to $O\left(H\right)$.


\section{Preliminaries}
\label{sec:perliminaries}
Given a set $\mathcal U$ and a positive integer $M\in\mathbb N^+$, we say that $\mathcal S$ is a $(\mathcal U, M)$-weighted stream if it contains a sequence of $\langle id, weight\rangle$ pairs.
Specifically:
\ifdefined \EXTENDED
$\mathcal S = \langle p_1,\ldots p_N\rangle,$
\fi
\ifdefined \NINEPAGES
$\mathcal S = \langle p_1,p_2,\ldots p_N\rangle,$
\fi
where $\forall i\in{1,\ldots, N}: p_i\in \mathcal U\times\set{1,\ldots M}$.
Given a packet $p_i=(d_i,w_i)$, we say that $d_i$ is $p_i$'s id while $w_i$ is its weight; $N$ is the \emph{stream length}, and $M$ is the \emph{maximal packet size}.
Notice that the same packet id may possibly appear multiple times in the stream, and each such occurrence may potentially be associated with a different weight.
Given a $(\mathcal U, M)$-weighted stream $\mathcal S$, we denote $v_x$, the \emph{volume} of id $x$, as the total weight of all packets with id $x$. That is:
$v_x\triangleq \sum_{\substack{i\in\set{1,\ldots,N}:\\d_i=x}}w_i.$
\normalsize
For a \emph{window size} $W\in\mathbb{N^+}$, we denote the \emph{window volume} of id $x$ as its total weight of packets with id $x$ within the last $W$~packets, that is:
$v^W_x\triangleq \sum_{\substack{i\in\set{N-W+1,\ldots,N}:\\d_i=x}}w_i.$
\normalsize
We seek algorithms that support the operations:

{\sc \textbf{ADD$\mathbf{(\langle x,w\rangle)}$}}: append a packet with identifier $x$ and weight $w$ to $\mathcal S$.

{\sc \textbf{Query$\mathbf{(x)}$}}: return an estimate  $\widehat{v_x}$ of $v_x$.

{\sc \textbf{WinQuery$\mathbf{(x)}$}}: return an estimate $\widehat{v^W_x}$ of $v^W_x$.

\noindent We now formally define the main problems in this work:

$\mathbf{(\epsilon,M)}$\textbf{-}{\sc \textbf{Volume Estimation}}: \query{} returns an estimation ($\widehat{v_x}$) that satisfies
$v_x \le \widehat{v_x} \le v_x + N\cdot M\cdot \epsilon.$
\normalsize

$\mathbf{(W,\epsilon,M)}$\textbf{-}{\sc \textbf{Volume Estimation}}: \winQuery{} returns an estimation ($\widehat{v^W_x}$) that satisfies $v^W_x \le \widehat{v^W_x} \le v^W_x + W\cdot M\cdot \epsilon.$
\normalsize

$\mathbf{(\theta,\epsilon,M)}$\textbf{-}{\sc \textbf{Approximate Weighted Heavy Hitters}}:\\ returns a set $H\subseteq\mathcal U$ such that:
\begin{align*}
\forall x\in \mathcal U:& (v_x > N\cdot M \cdot \theta\implies x\in H) \ \wedge\\ &(v_x < N\cdot M \cdot (\theta-\epsilon)\implies x\notin H).
\end{align*}
\normalsize
$\mathbf{(W, \theta,\epsilon,M)}$\textbf{-}{\sc \textbf{Approximate Weighted Heavy Hitters}}: returns a set $H\subseteq\mathcal U$ such that
\begin{align*}
\forall x\in \mathcal U:& (v^W_x > W\cdot M \cdot \theta\implies x\in H) \ \wedge\\ &(v^W_x < W\cdot M \cdot (\theta-\epsilon)\implies x\notin H).
\end{align*}
\normalsize

Our heavy hitter definitions are asymmetric.
That is, they require that flows whose frequency is above the threshold of $N\cdot M\cdot \theta$ (or $W\cdot M\cdot \theta$) are included in the list, but flows whose volume is \emph{slightly} less than the threshold can be either included or excluded from the list.
This relaxation is necessary as it enables reducing the required amount of space to sub linear.
Let us emphasize that the identities of the heavy hitter flows are not known in advance.
Hence, it is impossible to a-priori allocate counters only to these flows.
%
%
%
The basic notations used in this work are listed in Table~\ref{tbl:notations}.

\begin{table}[t]
	\centering
	\scriptsize
	\begin{tabular}{|c|l|}
		
		\hline
		Symbol & Meaning \tabularnewline
		\hline
		$S$ & stream \tabularnewline
		\hline
		$N$ & number of elements in the stream \tabularnewline
		\hline
		$M$ & maximal value of an element in the stream \tabularnewline
		\hline
		$W$ & window size \tabularnewline
		\hline
		$\mathcal U$ & the universe of elements \tabularnewline
		\hline
		$[r]$ & the set $\set{0,1,...,r-1}$ \tabularnewline
		\hline
		$\gamma$ & \rss{} performance parameter. \tabularnewline
		\hline
		$v_x$ & the volume of an element $x$ in $S$ \tabularnewline
		\hline
		$\widehat{v_x}$ & an estimation of $v_x$ \tabularnewline
		\hline
		$v^W_x$ & the volume of element $x$ in the last $W$ elements of $S$
		 \tabularnewline
		\hline
		$\widehat{v^W_x}$ & an estimation of $v^W_x$ \tabularnewline
		\hline
		$\epsilon$ & estimation accuracy parameter \tabularnewline
		\hline
		$\theta$ & heavy hitters threshold parameter \tabularnewline
		\hline
	\end{tabular}
	\caption{List of Symbols}
	\label{tbl:notations}
	\vspace{-2em}
\end{table}
\normalsize

\section{Frequent items Algorithm with a Semi-structured Table  (FAST)}
\label{sec:RSS}
\begin{figure}[t]
	\centering
\ifdefined\EXTENDED
	\includegraphics[width=\columnwidth]{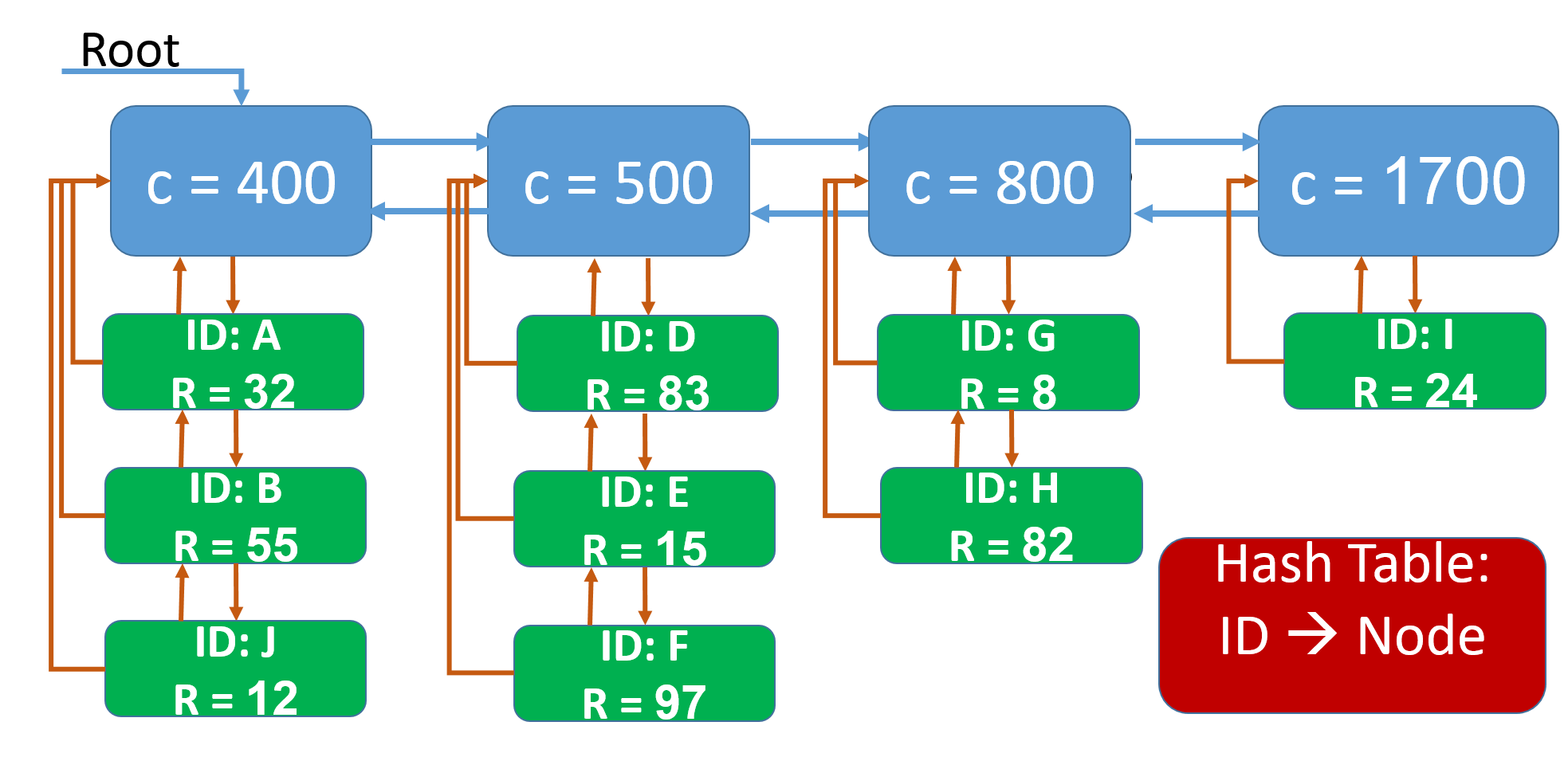}
\else
    \includegraphics[width=0.85\columnwidth]{RSS.png}
\fi
	\caption{An example of how \rss{} utilizes the SOS structure. Here, flows are partially ordered according to the third digit (100's), and each flow maintains its own remainder; e.g., the estimated volume of $D$ is $\widehat{v_D}=583$.
		}
	\label{fig:rss}
\end{figure}
In this section, we present \emph{\RSS{} (\rss{})}, a novel algorithm that achieves constant time weighted updates. FAST uses a data structure called \emph{Semi Ordered Summary (SOS)}, which maintains flow entries in a semi ordered manner.
That is, similarly to previous works, SOS groups flows according to their volume, each of which is called a \emph{volume group}.
The volume groups are maintained in an ordered list.
Each volume group is associated with a value $C$ that determines the volume of its nodes.
Unlike existing data structures, counters within each volume group are kept unordered.

Unlike previous works, the grouping is done at coarse granularity.
Each node (inside a group) includes a variable called \emph{Remainder} (denoted $R$).
The volume estimate of a flow is $C+R$ where $R$ is the remainder of its volume node and $C$ is the value of its volume group.

This semi-ordered structure is unique to SOS and enables it to serve weighted updates in $O(1)$.
Volume queries are satisfied in constant time using a separate aggregate hash table which maps between each flow identifier and its SOS node.
FAST then uses SOS to find a near-minimum flow when needed.

Figure~\ref{fig:rss} provides an intuitive example for the case $M = 1,000$.
Here, the volume of an item is calculated by both its group counter ($C$) and the item's remainder ($R$), e.g., the volume of A is $400+32 = 432$.
Flows are partially ordered according to their third digit, i.e., in multiples of $100$, or $M/10$.
Within a specific group, however, items are unordered, e.g., A, B and J are unordered but all appear before items with volume of at least $500$.
As the number of lists to skip prior to an addition is $O(1)$, the update complexity is also $O(1)$.
\ifdefined\EXTENDED
Specifically, we need to traverse at most $10$ linked lists when updating an item.
\fi

Intuitively, flows are only ordered according to volume groups and if we make sure that the maximal weight can only advance a flow a constant number of flow groups then SOS operates in constant time.
Alas, keeping the flows only partially ordered increases the error.
We compensate for such an increase by requiring a larger number of SOS entries compared to previously suggested fully ordered structures.
The main challenge in realizing this idea is to analyze the accuracy impact and provide strong estimation guarantees. 

\subsection{FAST - Accurate Description}
\rss{} employs $\ceil{\frac{1+\gamma}{\epsilon}}$ counters, for some non-negative constant $\gamma \ge 0$.
$\gamma$ determines how ordered SOS is: for $\gamma = 0$, we get full order, while for $\gamma>0$, it is only ordered up to $M \cdot \gamma/2$ (all flows that fall into the same volume group are unordered, and each group holds a range of $M \cdot \gamma/2$ values).
The runtime is, however, $O(1/\gamma)$ and is therefore constant for any fixed $\gamma$.
We note that an $\Omega\parentheses{\oneOverE}$ counters lower bound is known~\cite{SpaceSavings}.
Thus, \rss{} is asymptotically optimal for constant $\gamma$.
The pseudo code of FAST appears in Algorithm~\ref{alg:RSS}.

\begin{algorithm}[t]
\scriptsize
\begin{align*}
\text{Initialization: }\qquad&C\gets \emptyset, \forall x:c_x \gets 0, r_x \gets 0,
\\&\stepLetter \gets \step, \nrCountersLetter{}\gets \nrCounters
\end{align*}
\begin{algorithmic}[1]
\Function{Add}{Item $x$, Weight $w$}
	\If {$x\in C$ \textbf{ or } $|C| < \nrCountersLetter{}$}\label{line:firstIf}
		\State $c_x \gets c_x + \floor{	\frac{r_x + w}{\stepLetter}}$\label{line:regularCounterUpdate}
		\State $r_x \gets (r_x + w) \mod \stepLetter$  \label{line:regularRemainderUpdate}
		\State $C \gets C \cup \{x\}$
	\Else
		\State $\text{Let } m \in \text{argmin}_{y\in C}(c_y) $ \Comment{arbitrary minimal item}\label{line:min}
		\vspace{0.1cm}
		\State $c_x \gets c_m + \floor{	\frac{\stepLetter - 1 + w}{\stepLetter}}$\label{line:takeoverCounterUpdate}
		\State $r_x \gets (\stepLetter - 1 + w) \mod \stepLetter$  			\label{line:takeoverRemainderUpdate}
		\State $C \gets C \setminus\{m\} \cup \{x\}$		\label{line:eviction}	
	\EndIf

\EndFunction
\Function{Query}{$x$}
\If {$x\in C$ \textbf{ or } $|C| < \nrCountersLetter{}$}
	\State\Return $r_x + \stepLetter\cdot c_x$\label{line:normalQuery}
\Else
	\State\Return $\stepLetter - 1 + \stepLetter\cdot\min_{y\in C} c_y$\label{line:noCounterQuery}
\EndIf
\EndFunction

\end{algorithmic}
\normalsize
\caption{\rss{} ($M,\epsilon,\gamma$)}
\label{alg:RSS}
\end{algorithm}

\subsection{FAST Analysis}
We start by a simple useful observation
\begin{observation}\label{obs:modulo}
	Let $a,b\in\mathbb N: a = b\cdot \floor{\frac{a}{b}} + \parentheses{a\mod b}$.
\end{observation}
For the analysis, we use the following notations: for every item $x\in\mathcal U$ and stream length $t$, we denote by $q_t(x)$ the value of $\mbox{\query}$ after seeing $t$ elements.
We slightly abuse the notation and refer to $t$ also as the \emph{time} at which the $t^{\mathrm{th}}$ element arrived, where time here is discrete.
We denote by $C_t$ the set of elements with an allocated counter at time $t$, by $r_{x,t}$ the value of $r_x$ and by $c_{x,t}$ the value of $c_x$.
Also, we denote the volume at time $t$ as $v_{x,t} \triangleq \sum_{\substack{i\in\set{1,\ldots,t}:\\d_i=x}} w_i$.
\ifdefined\NINEPAGES
All missing proofs appear in Appendix~\ref{sec:missing-proofs}.
\fi

We now show that \rss{} has a one-sided error.
\begin{lemma}\label{lem:lowerBound}
For any $t\in\mathbb N$, after seeing any  $(\mathcal U, M)$-weighted stream $\mathcal S$ of length $t$,  for any $x\in\mathcal U:v_x \le \widehat{v_x}.$
\end{lemma}

\ifdefined\EXTENDED
\begin{proof}
We prove $v_{x,t}\le q_t(x)$ by induction over $t$. \\
\textbf{Basis:} $t = 0$. Here, we have $v_{x,t} = 0 = q_t(x)$. \\
\textbf{Hypothesis:} $v_{x,t-1}\le q_{t-1}(x)$\\
\textbf{Step:} $\angle{x_t,w_t}$ arrives at time $t$. By case~analysis:

Consider the case where the queried item $x$ is not the arriving one (i.e., $x\neq x_t$).
In this case, we have $v_{x,t} = v_{x,t-1}$.
If $x\in C_{t-1}$ but was evicted (Line~\ref{line:eviction}) then $c_x\in \text{argmin}_{y\in C_{t-1}}(c_{y,t-1})$. This means that:
\begin{multline*}
q_{t-1}(x)= r_{x,t-1}+\stepLetter\cdot\text{argmin}_{y\in C_{t-1}}(c_{y,t-1})\\
\le \stepLetter-1+\stepLetter\cdot\text{argmin}_{y\in C_t}(c_{y,t}) = q_t(x),
\end{multline*}
\normalsize
where the last equation follows from the query for $x\notin C_t$ (Line~\ref{line:noCounterQuery}).
Next, if $x\in C_{t-1}$ and $x\in C_t$, its estimated volume is determined by Line~\ref{line:normalQuery} and we get $q_t(x)=q_{t-1}(x) \ge v_{x,t-1}=v_{x,t}$.
If $x\notin C_{t-1}$ then $x\notin C_{t}$, so the values of $q_t(x),q_{t-1}(x)$ are determined by line~\ref{line:noCounterQuery}.
Since the value of $\min_{y\in C} c_y$ can only increase over time, we have $q_t(x)\ge q_{t-1}(x)\ge v_{x,t}$ and the claim holds.

On the other hand, assume that we are queried about the last item, i.e., $x=x_t$.
In this case, we get $v_{x,t} = v_{x,t-1}+w_t$.
We consider the following cases:
First, if $x\in C_{t-1}$, then $q_t(x) = q_{t-1}(x) + w_t$.
Using the hypothesis, we conclude that $v_{x,t}=v_{x,t-1}+w_t\le q_{t-1}(x)+w_t = q_t(x)$ as required.
Next, if $|C_{t-1}| < \nrCountersLetter$, we also have $q_t(x) = q_{t-1}(x) + w_t$ and the above analysis holds.
Finally, if $x\notin C_{t-1}$ and $|C_{t-1}|=\nrCountersLetter$, then
\footnotesize
\begin{align}
q_{t-1}(x) = s - 1 + s\cdot \min_{y\in C_{t-1}}c_{y,t-1}.\label{eq:qt1query}
\end{align}
\normalsize
On the other hand, when $x$ arrives, the condition of Line~\ref{line:firstIf} was not satisfied, and thus
\footnotesize
\begin{align*}
q_t(x) &= r_{x,t}+s\cdot c_{x,t} =(\stepLetter-1+w)\mod \stepLetter \\&\qquad+ s\cdotpa{ \min_{y\in C_{t-1}}c_{y,t-1} + \floor{	\frac{\stepLetter - 1 + w}{\stepLetter}}}\\
_{\parentheses{\tiny \text{Observation~\ref{obs:modulo}}}}&= s\cdot \min_{y\in C_{t-1}}c_{y,t-1} + \stepLetter - 1 + w\\
_{\eqref{eq:qt1query}}&=q_{t-1}(x) + w\\
_{\parentheses{\substack{\tiny \text{induction}\\\text{hypothesis}}}}&\ge v_{x,t-1} + w = v_{x,t}.\qquad\qquad\qquad\qquad\qed
\end{align*}
\normalsize
\end{proof}
\fi

We continue by showing that \rss{} is accurate if there are only a few distinct items.
\begin{lemma}\label{lem:exactCount}
	If the stream contains at most $\nrCounters$ distinct elements then \rss{} provides an exact estimation of an items volume upon query.
\end{lemma}
\ifdefined\EXTENDED
\begin{proof}
	Since $|C|\le\nrCountersLetter$, we get that the conditions in Line~\ref{line:firstIf} and Line~\ref{line:normalQuery} are always satisfied. Before the queried element $x$ first appeared, we have $r_x=c_x=0$ and thus \query$=0$. Once $x$ appears once, it gets a counter and upon every arrival with value $w$, the estimation for $x$ exactly increases by $w$, since $x$ never gets evicted (which can only happen in Line~\ref{line:min}).
\end{proof}
\fi

We now analyze the sum of counters in $C$.
\begin{lemma}\label{lem:sum-of-counters}
	For any $t\in\mathbb N$, after seeing any  $(\mathcal U, M)$-weighted stream $\mathcal S$ of length $t$,  FAST satisfies:\\ \centering{
	$\sum_{x\in C_t} \mbox{\query} \le t\cdot{ M\cdot (1+\gamma/2)}.$}
\end{lemma}

\ifdefined\EXTENDED
\begin{proof}
	We prove the claim by induction on the stream length~$t$.
 \textbf{Basis:} $t = 0$.\\
		In this case, all counters have value of $0$ and thus  \\$\sum_{x\in C_t} q_t(x) = 0 = t\cdotpa{ M\cdot (1+\gamma/2)}$.\\
\textbf{Hypothesis:} \footnotesize$\sum_{x\in C_{t-1}} q_{t-1}(x) \le (t-1)\cdot{ M\cdot (1+\gamma/2)}$\normalsize.\\
\textbf{Step:} $\angle{x_t,w_t}$ arrives at time $t$.
		We consider the following cases:
		\begin{enumerate}
			\item $x\in C_{t-1}$ or $|C_{t-1}|<\nrCounters$. In this case, the condition in Line~\ref{line:firstIf} is satisfied and thus $c_{x,t} = c_{x,t-1} + \floor{	\frac{r_{x,t-1} + w}{\stepLetter}}$ (Line~\ref{line:regularCounterUpdate}) and $r_{x,t} = (r_{x,t-1} + {w}) \mod {\stepLetter}$ (Line~\ref{line:regularRemainderUpdate}). By Observation~\ref{obs:modulo} we get
			\scriptsize
			\begin{align}
			\hspace{-0.0cm} q_t(x) &=_{\parentheses{\substack{\text{by line}\\\ref{line:normalQuery}}}} r_{x,t} + \stepLetter\cdot c_{x,t} \notag\\
			&= c_{x,t-1} + \floor{	\frac{r_{x,t-1} + w}{\stepLetter}} + (r_{x,t-1} + {w}) \mod {\stepLetter}\notag\\
			&\hspace{-0.0cm}= w + c_{x,t-1} + r_{x,t-1} = q_{t-1}(x) + w\label{eq:qt-val}.
			\end{align}
			\normalsize
			Since the value of a query for every $y\in C_t\setminus\{x\}$ remains unchanged, we get that
			\scriptsize
			\begin{align*}
			\sum_{y\in C_t} q_t(y) &= q_t(x) + \sum_{\substack{y\in C_{t-1}\\ y\neq x}} q_{t-1}(y) \\
			_{(\text{by }\eqref{eq:qt-val})} &= w + q_{t-1}(x) + \sum_{\substack{y\in C_{t-1}\\ y\neq x}} q_{t-1}(y)\\
			&= w + \sum_{\substack{y\in C_{t-1}}} q_{t-1}(y)\\
			_{\parentheses{\substack{\tiny \text{induction}\\\text{hypothesis}}}}&\le w + (t-1)\cdotpa{ M\cdot (1+\gamma/2)}\\
			&\le M + (t-1)\cdotpa{ M\cdot (1+\gamma/2)}\\
			_{\parentheses{\gamma\ge 0}}&\le t\cdotpa{ M\cdot (1+\gamma/2)}.
			\end{align*}
			\normalsize
			\item $x\notin C_{t-1}$ and $|C_{t-1}|=\nrCounters$. In this case, the condition of Line~\ref{line:firstIf} is false and therefore $c_{x,t} = c_{m,t-1} + \floor{	\frac{\stepLetter - 1 + w}{\stepLetter}}$~(Line~\ref{line:takeoverCounterUpdate}) and $r_{x,t} \gets (\stepLetter - 1 + w) \mod \stepLetter$~(Line~\ref{line:takeoverRemainderUpdate}).
			From Observation~\ref{obs:modulo} we get that
			\footnotesize
			\begin{align}
			\hspace{-0.0cm} q_t(x) &=_{\parentheses{\substack{\text{by Line}\\\ref{line:normalQuery}}}} r_{x,t} + \stepLetter\cdot c_{x,t} \notag\\
			&= c_{m,t-1} + \floor{	\frac{\stepLetter - 1 + w}{\stepLetter}} + (\stepLetter - 1 + {w}) \mod {\stepLetter}\notag\\
			&\hspace{-0.0cm}= w + c_{m,t-1} + \stepLetter - 1\notag\\
			& = q_{t-1}(m) - r_{m,t-1} + \floor{\frac{M\gamma}{2}} + w\notag\\
			&\le q_{t-1}(m) + \floor{\frac{M\gamma}{2}} + w\label{eq:qt-val}.
			\end{align}
			\normalsize
			As before, the value of a query for every $y\in C_t\setminus\{x\}$ is unchanged, and since $C_{t-1}\setminus C_t = \{m\}$,
			\scriptsize
			\begin{align*}
			\sum_{y\in C_t} q_t(y) &= q_t(x) - q_{t-1}(m) + \sum_{\substack{y\in C_{t-1}}} q_{t-1}(y)  \\
			_{(\text{by }\eqref{eq:qt-val})} &\le \floor{\frac{M\gamma}{2}} + w + \sum_{\substack{y\in C_{t-1}}} q_{t-1}(y) \\
			_{\parentheses{\substack{\tiny \text{induction}\\\text{hypothesis}}}}&\le \floor{\frac{M\gamma}{2}} + w + (t-1)\cdotpa{ M\cdot (1+\gamma/2)}\\
			&\le \floor{\frac{M\gamma}{2}} + M + (t-1)\cdotpa{ M\cdot (1+\gamma/2)}\\
			_{\parentheses{\gamma\ge 0}}&\le t\cdotpa{ M\cdot (1+\gamma/2)}.\qquad\qquad\qquad\qquad\qed
			\end{align*}
			\normalsize
		\end{enumerate}
\end{proof}
\fi

Next, we show a bound on  \rss{}'s estimation~error.
\begin{lemma}\label{lem:upperBound}
	For any $t\in\mathbb N$, after seeing any  $(\mathcal U, M)$-weighted stream $\mathcal S$ of length $t$, for any $x\in\mathcal U:\widehat{v_x}\le v_x+ t\cdot M\cdot\eps.$
\end{lemma}
\ifdefined\EXTENDED
\begin{proof}
First, consider the case where the stream contains at most $\nrCounters$ distinct elements.
By Lemma~\ref{lem:exactCount}, $\widehat{v_x}\le v_x$ and the claim holds.
Otherwise, we have seen more than $\nrCounters$ distinct elements, and specifically
\footnotesize
\begin{align}
t > \nrCounters\label{eq:streamLen}.
\end{align}
\normalsize
From Lemma~\ref{lem:sum-of-counters}, it follows that
\ifdefined \NINEPAGES
\small
\fi
\scriptsize
\begin{align}
\min_{y\in C_t} Query(y) \le \frac{t\cdot M \cdotpa{1+\gamma/2}}{\nrCounters}\le\frac{t\cdot M \cdot\eps\cdotpa{1+\gamma/2}}{1+\gamma}.\label{eq:minBound}
\end{align}
\normalsize
Notice that $\forall x\in C_t$,
\ifdefined \EXTENDED
the value of
\fi
\query{} is determined in Line~\ref{line:normalQuery}; that is, $q_{t}(x) = r_{x,t} + s\cdot c_{x,t}$. Next, observe that an item's remainder value is bounded by $\stepLetter-1$ (Line~\ref{line:regularRemainderUpdate} and Line~\ref{line:takeoverRemainderUpdate}). Thus,
\footnotesize
\begin{align}
\forall x,y\in C_t:q_t(x)\ge \stepLetter + q_t(y)\implies c_{x,t} > c_{y,t}.\label{eq:minCondition}
\end{align}
\normalsize
By choosing $y\in\arg\min_{y\in C_t}q_t(y)$, we get that if $v_{x,t}\ge q_t(y) + \stepLetter$, then $q_t(x)\ge q_t(y) + \stepLetter$ and thus $c_{x,t} > c_{y,t}$. Next, we show that if $v_{x,t} \ge t\cdot M \cdot \eps$, then $c_x > \min_{y\in C_t} c_y$ and thus $x$ will never be the ``victim'' in Line~\ref{line:min}:
\scriptsize
\begin{align*}
q_t(x)&\ge v_{x,t}\ge t\cdot M \cdot \eps = t\cdot M \cdot \eps \cdot\frac{1+\gamma/2}{1+\gamma} + M\gamma/2\cdot\frac{t}{\frac{\slack}{\eps}}\\
_\eqref{eq:minBound} &\ge q_t(y) + M\gamma/2\cdot\frac{t}{\frac{\slack}{\eps}}\\
_\eqref{eq:streamLen} &> q_t(y) + M\gamma/2 .
\end{align*}
\normalsize
Next, since $q_t(x)$ and $q_t(y)$ are integers, it follows that
$$q_t(x) \ge q_t(y) + \step = q_t(y)+\stepLetter.$$
Finally, we apply \eqref{eq:minCondition} to conclude that once $x$ arrives with a cumulative volume of $t\cdot M\cdot\eps$, it will never be evicted (Line~\ref{line:min}) and from that moment on its volume will be measured exactly.\qquad
\end{proof}
\fi

Next, we prove a bound on the run time of \rss{}.
\begin{lemma}\label{lem:runtime}
	let $\gamma > 0$, \rss{} adds in $O\parentheses{\oneOverG}$ time.
\end{lemma}

\ifdefined\EXTENDED
\begin{proof}
As mentioned before, \rss{} utilizes the SOS data structure that answers queries in $O(1)$.
Updates are a bit more complex as we need to handle weights and thus may be required to move the flow more than once, upon a counter increase.
Whenever we wish to increase the value of a counter (Line~\ref{line:regularCounterUpdate} and Line~\ref{line:takeoverCounterUpdate}), we need to remove the item from its current group and place it in a group that has the increased $c$ value.
This means that for increasing a counter by $n\in\mathbb N$, we have to traverse at most $n$ groups until we find the correct location.
Since the remainder value is at most $\stepLetter-1$ (Line~\ref{line:regularRemainderUpdate} and Line~\ref{line:takeoverRemainderUpdate}), we get that at any time point, a counter is increased by no more than  $\floor{	\frac{\stepLetter - 1 + w}{\stepLetter}}$ (Line~\ref{line:regularCounterUpdate} and Line~\ref{line:takeoverCounterUpdate}).
Finally, since $\stepLetter = \step$, we get that the counter increase is bounded by
	\ifdefined \NINEPAGES
	\small
	\begin{align*}
	\floor{	\frac{\ensuremath{\floor{M\cdot\gamma/2 + 1}} - 1 + w}{\ensuremath{\floor{M\cdot\gamma/2 + 1}}}}< 1 + \frac{w}{M\gamma/2} \le 1 + \frac{2}{\gamma} = O\parentheses{\oneOverG}.\ 
	\normalsize
	\end{align*}
	\fi
	\ifdefined \EXTENDED
    \scriptsize
	\begin{align*}
	\floor{	\frac{\step - 1 + w}{\step}}< 1 + \frac{w}{\frac{M\gamma}{2}} \le 1 + \frac{2}{\gamma} = O\parentheses{\oneOverG}.\quad\qed
	\end{align*}
    \normalsize
	\fi
\end{proof}
\fi

Next, we combine Lemma~\ref{lem:lowerBound}, Lemma~\ref{lem:upperBound} and Lemma~\ref{lem:runtime} to conclude the correctness of the \rss{} algorithm.
\ifdefined \NINEPAGES
\begin{theorem}
	For any constant $\gamma>0$, when allocated $\nrCountersLetter{}\triangleq \nrCounters$ counters, \rss{} operates in constant time and solves the \frqEst{} problem.
\end{theorem}
\fi
\ifdefined \EXTENDED
\begin{theorem}
	For any fixed $\gamma>0$, when allocated with $\nrCountersLetter{}\triangleq \nrCounters$ counters, \rss{} performs updates and queries in constant time, and solves the \frqEst{} problem.
\end{theorem}
\fi

Finally, \rss{} also solves the heavy hitters problem:
\begin{theorem}
	For any fixed $\gamma>0$, when allocated with $\nrCountersLetter{}\triangleq$\normalfont $\nrCounters$ counters, by returning $\{x\in \mathcal U\mid \widehat{v_x}\ge N\cdot M\cdot \theta\}$, \rss{} solves the \heavyHitters{} problem.
\end{theorem}

\section{Windowed \rss{} (W\rss{})}
\label{sec:sliding-window}
We now present \emph{Windowed \RSS{} (W\rss)}, an efficient algorithm for the ${(W,\epsilon,M)}$\textbf{-}{\sc {Volume Estimation}} and ${(W, \theta,\epsilon,M)}$\textbf{-}{\sc {Weighted Heavy Hitters}} problems.

\begin{figure}[t]
\centering
\includegraphics[width=\linewidth]{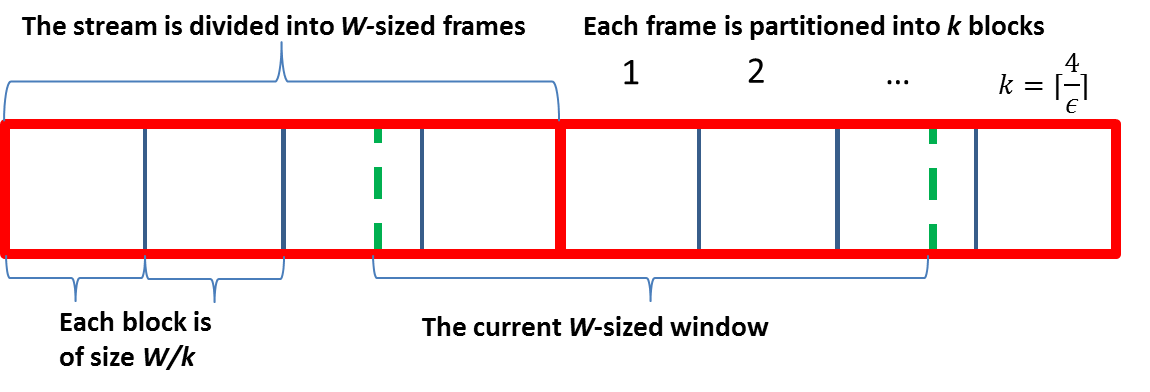}
\caption{The stream is divided into intervals of size $\window$ called \emph{frames} and each frame is partitioned into $\numBlocks$ equal-sized \emph{blocks}. The window of interest is also of size $\window$, and overlaps with at most $2$ frames and $\numBlocks+1$ blocks.}
\label{fig:window-counting}
\end{figure}

We partition the stream into consecutive sequences of size $W$ called \emph{frames}.
Each frame is further divided into $\numBlocks\triangleq\ceil{\frac{4}{\eps}}$ \emph{blocks}, each of size $\blockSize$, which we assume is an integer for simplicity.
Figure~\ref{fig:window-counting} illustrates the
\ifdefined \EXTENDED
\fi
setting.

W\rss{} uses a \rss{} instance $y$ to estimate the volume of each flow within the current frame.
Once a frame ends (the stream length is divisible by $W$), we ``flush'' the instance, i.e., reset all counters and remainders to $0$.
Yet, we do not ``forget'' all information in a flush, as high volume flows are stored in a dedicated data structure.
Specifically, we say that an element $x$ \emph{overflowed} at time $t$ if $\floor{\frac{q_{x,t}}{MW/k}} > \floor{\frac{q_{x,t-1}}{MW/k}}$.
We use a queue of queues structure $b$ to keep track of which elements have overflowed in each block.
That is, each node of the main queue represents a block and contains a queue of all elements that overflowed in its block.
Particularly, the secondary queues maintain the ids of overflowing elements.
Once a block ends, we remove the oldest block's node (queue) from the main queue, and initialize a new queue for the starting block.
Finally, we answer queries about the window volume of an item $x$ by multiplying its overflows count by $MW/k$, adding the residual count from $y$ (i.e., the part that is not recorded in $b$), plus $2MW/k$ to ensure an overestimation.

For $O(1)$ time queries, we also maintain a hash table $B$ that tracks the overflow count for each item.
That is, for each element $x$, $B[x]$ contains the number of times $x$ is recorded in $b$.
Since multiple items may overflow in the same block, we cannot update $B$ once a block ends in constant time.
We address this issue by \emph{deamortizing} $B$'s update, and on each arrival we remove a \emph{single} item from the queue of the oldest block (if such exists).
The pseudo code of W\rss{} appears in Algorithm~\ref{alg:WRSS} and a list containing its variables description appears in Table~\ref{tbl:sliding-window-vars}.
An efficient implementation of the queue of queues $b$ is described in~\cite{WCSS}.
\begin{table}[t]
\scriptsize
\begin{tabular}{|c|l|}
\hline
$k$ & A constant $k \triangleq \ceil{4 / \varepsilon}$
\tabularnewline
\hline
$\SSSInstance$ & A \rss{} instance using $\numBlocks(1+\gamma)$ counters. \tabularnewline
\hline
$\queueOfOverflows$ & A queue of $\numBlocks+1$ queues. \\&An efficient implementation appears in~\cite{WCSS}. \tabularnewline
\hline
$\sumOfOverflows$ & The histogram of $\queueOfOverflows$, implemented using a hash table. \tabularnewline
\hline
$\frameOffset$ & The offset within the current frame. \tabularnewline
\hline
\end{tabular}
\normalsize
\caption{Variables used by the W\rss{} algorithm.}
\label{tbl:sliding-window-vars}
\end{table}



\begin{algorithm}[t!]
\caption{WFAST ($W, M, \gamma$)}\label{alg:WRSS}
\scriptsize
\begin{algorithmic}[1]
\ifdefined \NINEPAGES
\State Initialization: $\SSSInstance\gets\RSS{}(M,1/k,\gamma), \frameOffset \gets 0, $\\
\hspace{1.38cm}$\sumOfOverflows\gets\mbox{Empt hash table}, $
$\queueOfOverflows \gets \mbox{Queue of $\numBlocks+1$ empty queues}$.
\fi
\ifdefined \EXTENDED
\Statex Initialization: $\SSSInstance\gets\rss{}(M,1/k,\gamma), $
\Statex $\frameOffset \gets 0, \sumOfOverflows\gets\mbox{Empty hash table}, $
\Statex
$\queueOfOverflows \gets \mbox{Queue of $\numBlocks+1$ empty queues}$.
\fi
\Function{add}{Item $x$, Weight $w$}
\State \inc\frameOffset $\mod W$ 
\If {$\frameOffset = 0$} \Comment new frame starts
	\State$y$.\Call {flush}{\null}\label{line:flush}
\EndIf
\If {$\frameOffset\mod\blockSize = 0$}
	\Comment new block
	\State$\queueOfOverflows$.\Call {pop}{\null}\label{line:pop}
	\State$\queueOfOverflows$.\Call {append}{new empty queue}
\EndIf
\If {$\queueOfOverflows$.tail is not empty} \Comment remove oldest item \label{line:empty-tail}
	\State $oldID \gets\queueOfOverflows$.tail.\Call {pop}{\null}\label{line:pop-from-b}
	\State \dec{\sumOfOverflows[oldID]}
	\If {$\sumOfOverflows[oldID]=0$}
		\State $\sumOfOverflows$.\Call{remove}{$oldID$}
	\EndIf
\EndIf
\State $prevOverflowCount \gets \floor{\frac{\SSSInstance\mbox{.\Call{query}{$x$}}}{MW/k}}$
\State$\SSSInstance.$\Call{add}{$x, w$}  \Comment add item
\If {$\floor{\frac{\SSSInstance\mbox{.\Call{query}{$x$}}}{MW/k}} > prevOverflowCount$} \Comment{overflow}\label{line:overflow}
	\State $\queueOfOverflows$.head.\Call{push}{$x$}\label{line:push}
	\If {\sumOfOverflows.\Call{contains}{x}}
	\State \inc{\sumOfOverflows[x]}
	\Else
	\State $\ \sumOfOverflows[x]\gets 1$ \Comment{adding $x$ to $\sumOfOverflows$}
	\EndIf
\EndIf
\EndFunction
\ifdefined \EXTENDED
\Statex
\fi
\Function{{\sc WinQuery}}{Item x}
\If {$\sumOfOverflows$.\Call{Contains}{$x$}}
\State\Return
\ifdefined \NINEPAGES
	 {$MW/k\cdot\parentheses{\sumOfOverflows[x]+2} + \parentheses{\SSSInstance.\Call{query}{x} \mod MW/k}$}\label{line:WRSSQuery}
\fi
\ifdefined \EXTENDED
{
	\scriptsize
	$MW/k\cdot\parentheses{\sumOfOverflows[x]+2} + \parentheses{\SSSInstance.\Call{query}{x} \mod MW/k}$
	\normalsize
}\label{line:WRSSQuery}
\fi
\Else \Comment {$x$ has no overflows}
	\State\Return $2MW/k$+y.\Call{query}{$x$}
\EndIf
\EndFunction
\end{algorithmic}
\normalsize
\end{algorithm}

%

\subsection{WFAST  Analysis}

We start by introducing several notations to be used in this section.
We mark the queried element by $x$, the current time by $\window+\frameOffset$, and assume that item $\window$ is the first element of the current frame.
For convenience, denote
\ifdefined \EXTENDED
$v_x(t_1,t_2)\triangleq \sum_{\substack{i\in\set{t_1,\ldots, t_2}:\\x_i = x}} w_i,$
\fi
\ifdefined \NINEPAGES
$v_x(t_1,t_2)\triangleq \sum_{\substack{i\in\set{t_1,\ldots, t_2}:\\x_i = x}} w_i,$
\fi
i.e., the volume of $x$ between $t_1$ and $t_2$.
The goal is then to approximate the window volume of $x$, which is defined as
\small
\ifdefined \EXTENDED
\scriptsize
\begin{align}\label{eq:goal}
v_x^w \triangleq v(\frameOffset+1,W+\frameOffset) =  v(\frameOffset+1,W-1) + v(W,W+\frameOffset),
\end{align}
\normalsize
\fi
\ifdefined \NINEPAGES
$
v_x^w \triangleq v(\frameOffset+1,W+\frameOffset) ,
$
\fi
\normalsize
i.e., the sum of weights in the timestamps within $\langle\frameOffset+1,\frameOffset+2,\ldots,\window+\frameOffset\rangle$ in which $x$ arrived.
\ifdefined \EXTENDED
We denote the value returned from $\SSSInstance\mbox{.\Call{query}{$x$}}$ after the $t$'th item was added by $\SSSInstance_t$.
Similarly, $\overflowIndicator_t$ represents whether $x$ arrived at time $t$ ($x = x_t$) \emph{and} overflowed, i.e., the condition of Line~\ref{line:overflow} was satisfied.
We assume that if $x$ is not allocated a \rss{} counter at time $t$, then $B[x]=0$, which allows us to consider Line~\ref{line:WRSSQuery} for queries.
For simplicity, mark $\sumOfOverflows[x]=0$ for $x \notin \sumOfOverflows$.

We proceed with a useful lemma
that bounds WFAST's error on arrivals happening before the flush (Line~\ref{line:flush}).

\begin{lemma}\label{lem:prevFrame}
Let $t_1,t_2\in\{\frameOffset+1,\ldots+\window-1\}$ be two timestamps within the previous frame,
then
$$
\floor{\frac{v_x(t_1,t_2)}{MW/k}}
\le\sum_{\tau=t_1}^{t_2} u_\tau
\le\ceil{\frac{v_x(t_1,t_2)}{MW/k}}.
$$
\end{lemma}
\begin{proof}
Since $\SSSInstance$, initialized with $\epsilon=\frac{1}{k}$, is flushed every $W$ elements (Line~\ref{line:flush}), any element $z$ that satisfies $\SSSInstance\mbox{.\Call{query}{$z$}} > MW/k$ is guaranteed not to lose its counter (see Lemma~\ref{lem:upperBound}'s proof for a similar analysis).
This means that once an item overflows, it never loses its counter.
Further, being an over-estimator, an element with a volume of $MW/k$ (or more) is guaranteed to have a counter since the minimal counter cannot exceed $MW/k$.
Thus, if $v_x(t_1,t_2)<MW/k$ then the claim holds, since after overflowing for the first time, an item has to arrive with a weight of $MW/k$ to overflow again.
On the other hand, if $v_x(t_1,t_2)>MW/k$ then $x$ may overflow for the first time before its volume reached $MW/k$, but from that point on only arrivals of $x$ increase the counter and can cause an overflow.
\end{proof}
We continue with proving the algorithm's correctness.
\fi
\ifdefined \NINEPAGES
We next state the main correctness theorem for W\rss.
\fi
\begin{theorem}\label{thm:window-algorithm-correctness}
Algorithm~\ref{alg:WRSS} solves the  ${(W,\epsilon,M)}${-}{\sc {Volume Estimation}} problem.
\end{theorem}
\ifdefined \NINEPAGES
Due to lack of space, the proof of the Theorem appears in the Appendix. 
\fi
\ifdefined \EXTENDED
\begin{proof}
We prove the theorem in two steps.
We first analyze the volume of $x$ within the previous frame, and then consider the current one.
We continue by bounding the error introduced by the deamortization and factor \rss{} being an approximation algorithm in the first place.
Finally, we add up the different error types and show that W\rss{} provides a decent approximation.

We start with the number of times $x$ has overflowed before the flush.
By applying Lemma~\ref{lem:prevFrame}, we get that
\scriptsize
\begin{align}
\floor{v_x(\frameOffset+1,W-1) \over MW/k}
\le\sum_{t=M+1}^{W-1} u_t
\le\ceil{v_x(\frameOffset+1,W-1) \over MW/k}.
\end{align}
\normalsize
We continue with analyzing the current frame, which started after $\SSSInstance$ was last flushed (Line~\ref{line:flush}).
As discussed above, an element whose volume is larger than $MW/k$ overflows and will not lose its counter in the flush,  thus:
\scriptsize
\begin{align}
\SSSInstance_{\window+\frameOffset}=MW/k\cdot\sum\nolimits_{t=\window}^{\window+\frameOffset}{\overflowIndicator_t}+\parentheses{\SSSInstance_{\window+\frameOffset} \mod MW/k}\label{eq:modulu}.
\end{align}
\normalsize
Notice that if $x$ does not have a counter, it did not overflow and the equation still holds.

Next, we consider the number of overflows recorded in $B[x]$ and the number of actual overflows.
We have deamortized (Line~\ref{line:pop-from-b}) the process of updating the overflow count (in $B$).
This means that $B[x]$ is not guaranteed to have the exact count of the number of times $x$ overflowed within the blocks overlapping with the current window.
Luckily, since $x$ cannot overflow twice in the same block, and specifically in the oldest block ($b.tail$), we get that we underestimate the number of overflows by at most one, and specifically:
\scriptsize
\begin{align}
\sum\nolimits_{t=\frameOffset+1}^{\window+\frameOffset}{\overflowIndicator_t}-1
\le\sumOfOverflows[x]
\le\sum\nolimits_{t=\frameOffset+1}^{\window+\frameOffset}{\overflowIndicator_t}.\label{eq:overflow-count}
\end{align}
\normalsize

Since $\SSSInstance$ is a \rss{} instance with parameters ($M,\frac{1}{\numBlocks},\gamma$), it solves the \frqEst{} problem, thus
\scriptsize
\begin{align}
v(W,W+\frameOffset)
\le \SSSInstance_{\window+\frameOffset}
\le v(W,W+\frameOffset)+MW/k.\label{eq:ss-approx}
\end{align}
\normalsize

When queried for $x$, the algorithm returns
\scriptsize
\begin{align*}
\xWindowFrequencyEstimator &= MW/k\cdot\parentheses{\sumOfOverflows[x]+2}+\parentheses{\SSSInstance_{\window+\frameOffset} \mod MW/k}\\
&=_{\eqref{eq:modulu}}
MW/k\cdot\parentheses{\sumOfOverflows[x]+2-\sum\nolimits_{t=W}^{\window+\frameOffset}{\overflowIndicator_t}} + \SSSInstance_{\window+\frameOffset}.
\end{align*}

\footnotesize

\ifdefined \EXTENDED
\begin{figure*}[t]
	
		\subfigure[SanJose14]{\includegraphics[width = \matrixCellWidth]
			{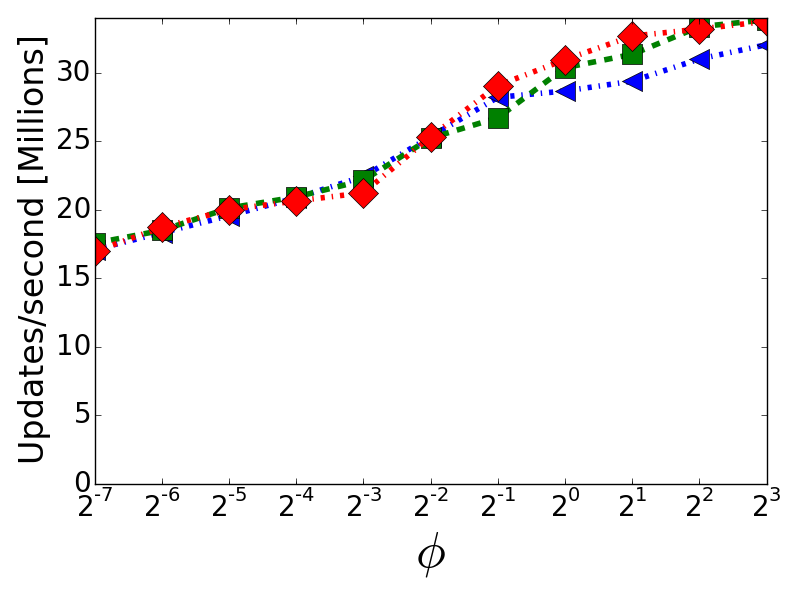}}	
		\subfigure[YouTube]{\includegraphics[width = \matrixCellWidth]
			{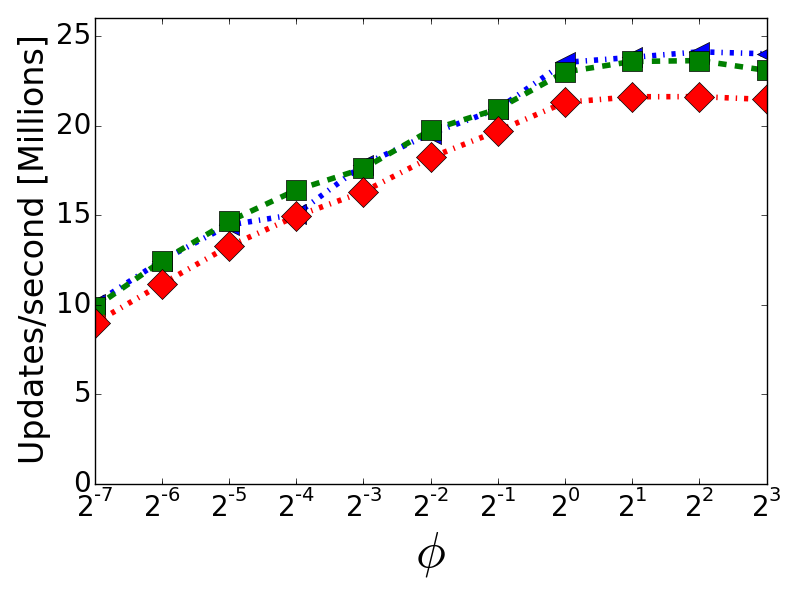}}	
		\subfigure[Chicago16]{\includegraphics[width = \matrixCellWidth]
			{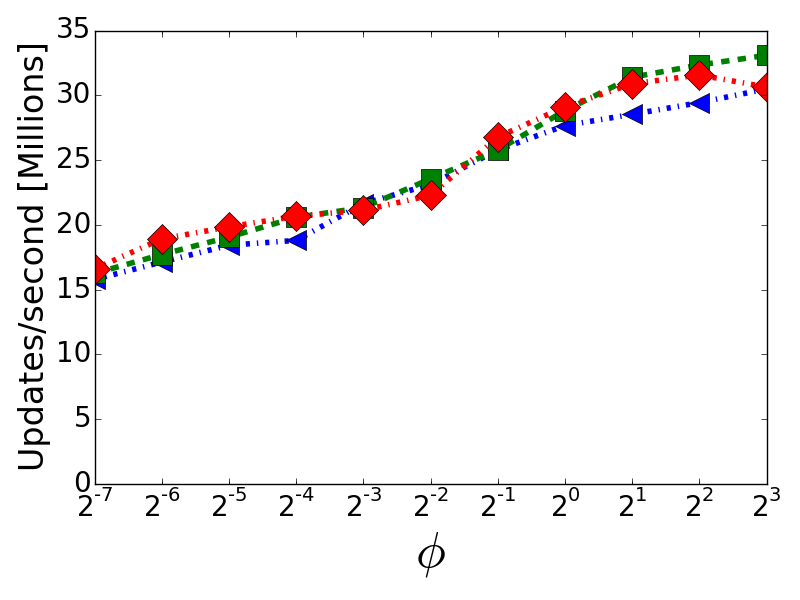}}
		\center{\includegraphics[width=0.8\columnwidth]{gammaGraphs/gammaLegend.png}}\\
		\subfigure[SanJose13]{\includegraphics[width = \matrixCellWidth]
			{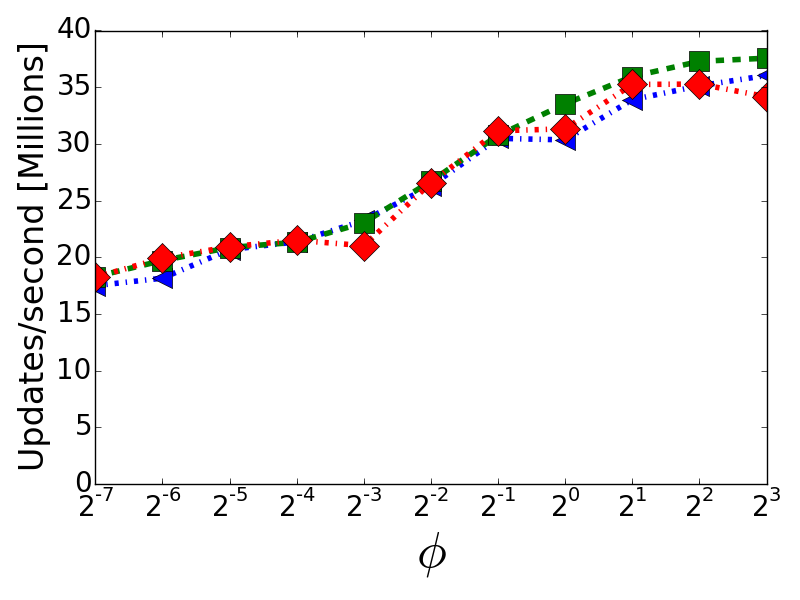}}
		\subfigure[DC1]{\includegraphics[width =  \matrixCellWidth]
			{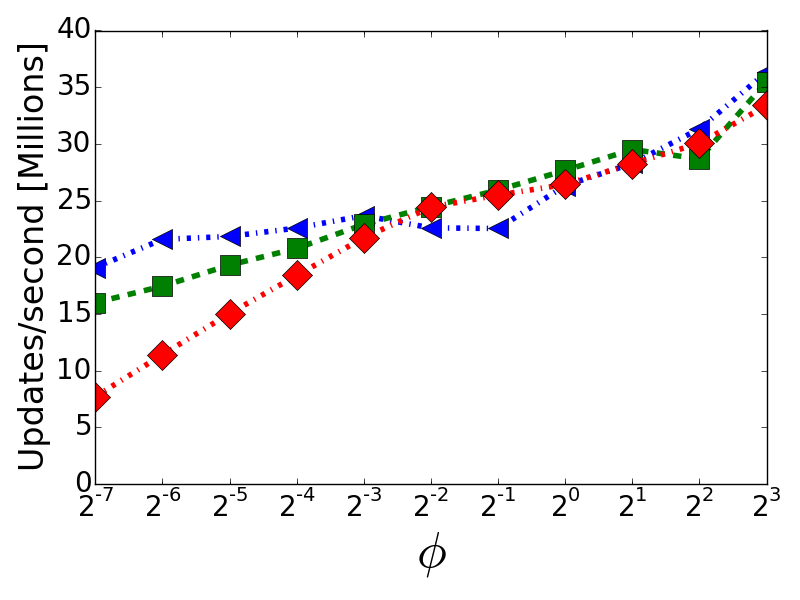}} 		
		\subfigure[Chicago15]{\includegraphics[width = \matrixCellWidth]
			{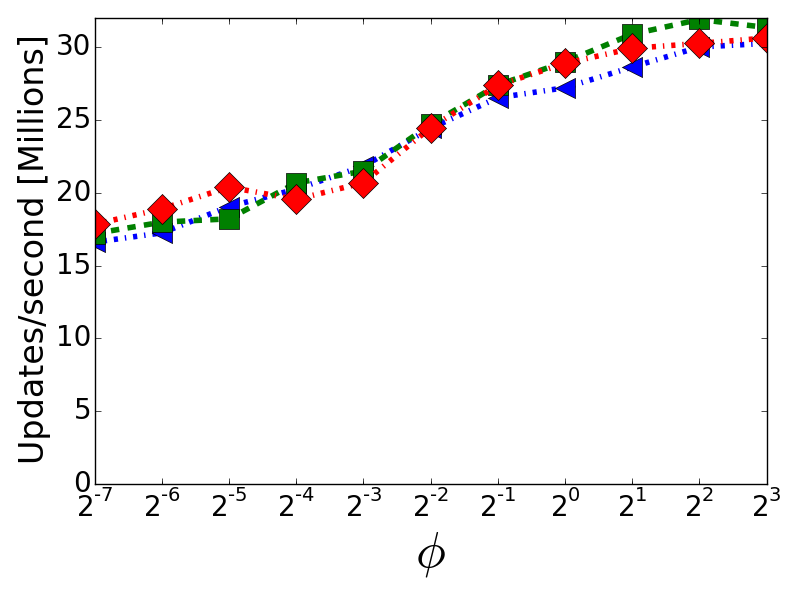}}
	\caption{	\label{fig:gamma}The effect of parameter $\gamma$ on operation speed for different error guarantees ($\epsilon$). $\gamma$ influences the space requirement as the algorithm is allocated with $\nrCounters$ counters.}
\end{figure*}
\fi

\begin{figure*}[t]
	\captionsetup{justification=centering}

		\subfigure[SanJose14]{\includegraphics[width = \matrixCellWidth]
			{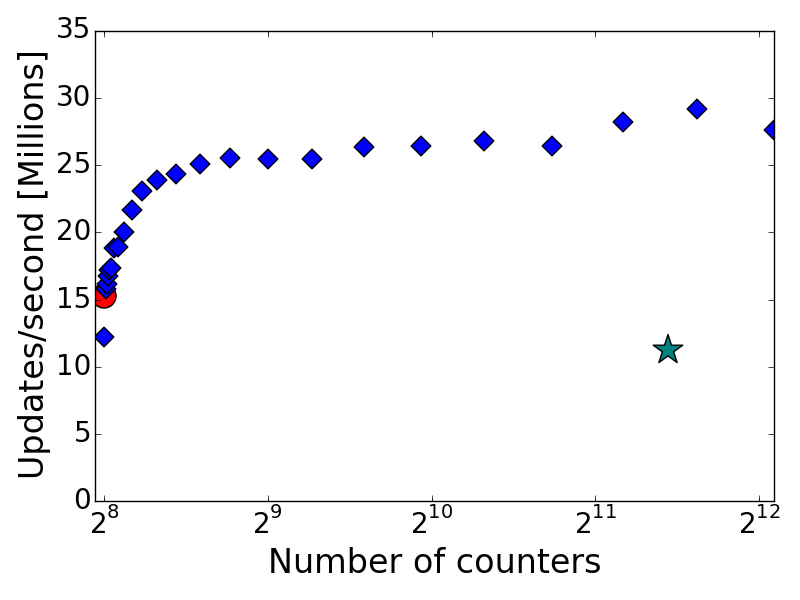}}
		\subfigure[YouTube]{\includegraphics[width = \matrixCellWidth]
			{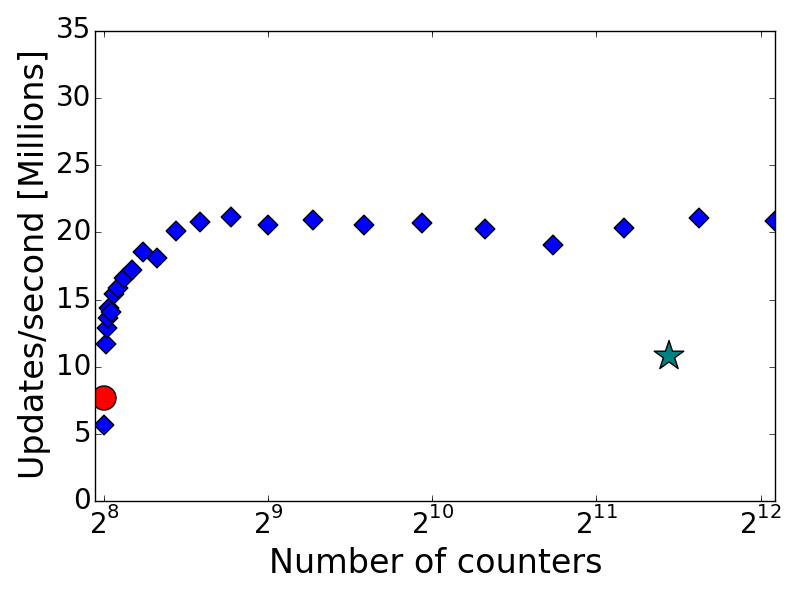}} 	
		\subfigure[Chicago16]{\includegraphics[width = \matrixCellWidth]
			{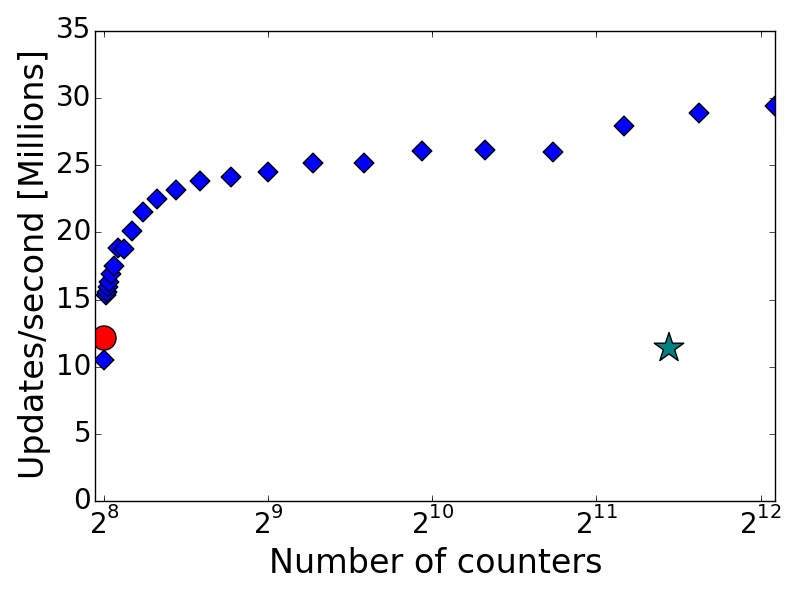}}
		\center{\includegraphics[width=0.8\columnwidth]{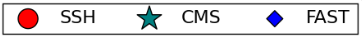}}\\
		\subfigure[SanJose13]{\includegraphics[width = \matrixCellWidth]
			{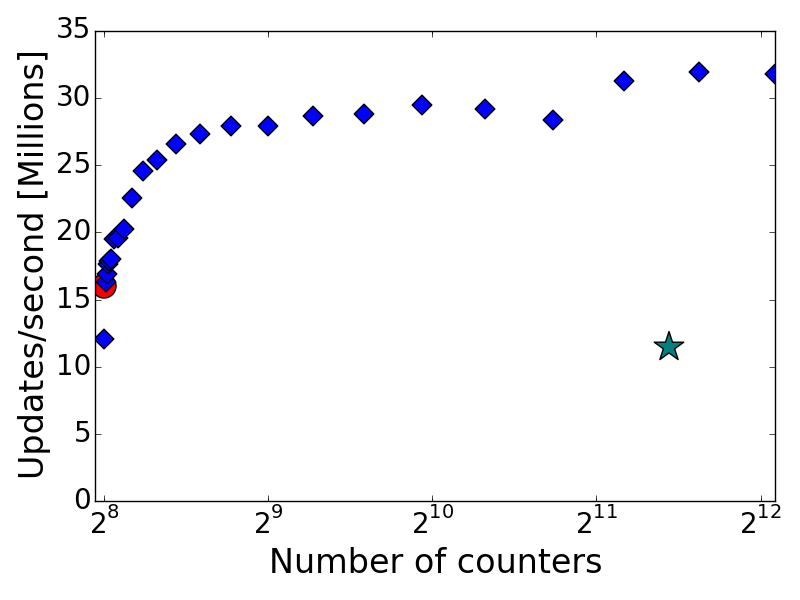}}
		\subfigure[DC1]{\includegraphics[width = \matrixCellWidth]
			{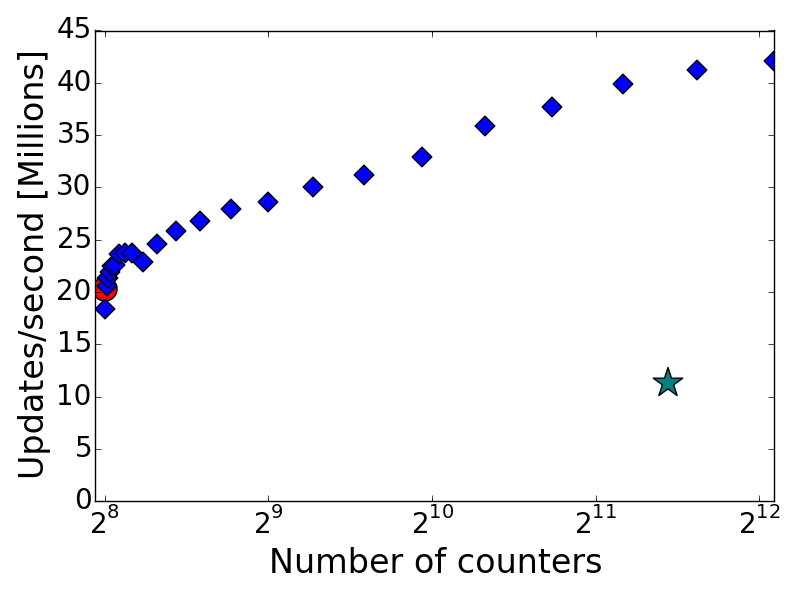}}  		
		\subfigure[Chicago15]{\includegraphics[width = \matrixCellWidth]
			{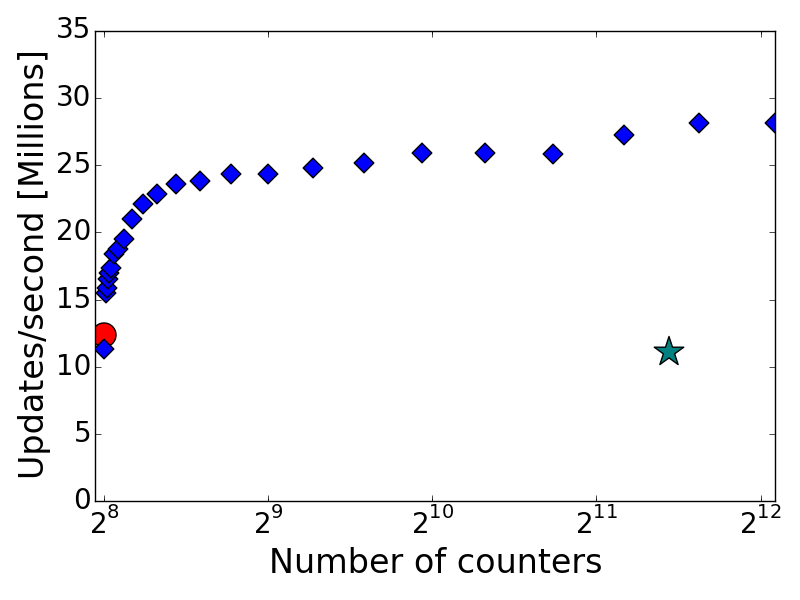}}
	\caption{\label{fig:tradeoff}Runtime comparison for a given error guarantee ($\epsilon=2^{-8}$). All algorithms provide the same guarantees and \rss{} uses different $\gamma$ values to show the speedup gained from allocating additional counters. }
\end{figure*}
%
%

\normalsize
We combine the above inequalities to bound the algorithms overestimation as follows:
\scriptsize
\begin{align*}
\xWindowFrequencyEstimator &= MW/k\cdot\parentheses{\sumOfOverflows[x]+2-\sum\nolimits_{t=W}^{\window+\frameOffset}{\overflowIndicator_t}} + \SSSInstance_{\window+\frameOffset}\\
&\le_{\eqref{eq:ss-approx}} MW/k\cdot\parentheses{\sumOfOverflows[x]+3-\sum\nolimits_{t=W}^{\window+\frameOffset}{\overflowIndicator_t}}+v(W,W+\frameOffset)\\
&\le_{\eqref{eq:overflow-count}}MW/k\cdot\parentheses{\sum\nolimits_{t=\frameOffset+1}^{\window+\frameOffset}{\overflowIndicator_t}+3-\sum\nolimits_{t=W}^{\window+\frameOffset}{\overflowIndicator_t}}+v(W,W+\frameOffset)\\
&=MW/k\cdotpa{\sum\nolimits_{t=\frameOffset+1}^{\window-1}{\overflowIndicator_t}+3}+v(W,W+\frameOffset)\\
&\le_{(Lemma~\ref{lem:prevFrame})}
MW/k\cdotpa{\ceil{\frac{v_x(\frameOffset+1,W-1)}{MW/k}}+3}+
\\&v(W,W+\frameOffset)\\
&\le_{\eqref{eq:goal}}v(\frameOffset+1,W+\frameOffset)+4MW/k \le  \xWindowFrequency+WM\eps.
\end{align*}
\normalsize
Similarly, we bound the query value from below:
\scriptsize
\begin{align*}
\xWindowFrequencyEstimator &= MW/k\cdot\parentheses{\sumOfOverflows[x]+2-\sum\nolimits_{t=W}^{\window+\frameOffset}{\overflowIndicator_t}} + \SSSInstance_{\window+\frameOffset}\\
&\ge_{\eqref{eq:ss-approx}}
MW/k\cdot\parentheses{\sumOfOverflows[x]+2-\sum\nolimits_{t=W}^{\window+\frameOffset}{\overflowIndicator_t}}+v(W,W+\frameOffset)\\
&\ge_{\eqref{eq:overflow-count}}MW/k\cdotpa{\sum\nolimits_{t=\frameOffset+1}^{\window+\frameOffset}{\overflowIndicator_t}+1-\sum\nolimits_{t=W}^{\window+\frameOffset}{\overflowIndicator_t}}+v(W,W+\frameOffset)\\
&=MW/k\cdotpa{\sum\nolimits_{t=\frameOffset+1}^{\window-1}{\overflowIndicator_t}+1}+v(W,W+\frameOffset)\\
&\ge_{(Lemma~\ref{lem:prevFrame})}
MW/k\cdotpa{\floor{\frac{v_x(\frameOffset+1,W-1)}{MW/k}}+1}+v(W,W+\frameOffset)\\
&\ge_{\eqref{eq:goal}}v(\frameOffset+1,W+\frameOffset)=
  \xWindowFrequency.
\end{align*}
\normalsize
Showing both bounds, we established that W\rss{} solves the ${(W,\epsilon,M)}${-}{\sc {Frequency Estimation}} problem.
\end{proof}
\else
\footnotesize

\normalsize
\fi
As a corollary, 
Algorithm~\ref{alg:WRSS} can 
find heavy hitters.
\begin{theorem}
\label{WFAST:HH}
By returning all items $x\in\mathcal U$ for which $\xWindowFrequencyEstimator\ge MW\theta$,
Algorithm~\ref{alg:WRSS} solves
\ifdefined \NINEPAGES
\small
\fi
\footnotesize
${(W, \theta,\epsilon,M)}${-}{\sc {Weighted Heavy~Hitters}}.\normalsize
\end{theorem}

\subsubsection*{W\rss{} runtime analysis:}
As listed in the pseudo code of W\rss{} (see Algorithm~\ref{alg:WRSS}) and the description above, processing new elements requires adding them to the \rss{} instance $\SSSInstance$, which takes $O(\oneOverG)$ time, and another $O(1)$ operations.
The query processing includes $O(1)$ operations and hash tables accesses.
For returning the heavy hitters, we go over all of the items with allocated counters in time $O(\frac{\slack}{\eps})$.
In summary, we get the following theorem:
\begin{theorem}
For any fixed $\gamma>0$, W\rss{} processes new elements and answers window-volume queries in constant time, while finding the window's weighted heavy hitters in $O(\oneOverE)$ time.
\end{theorem}

\begin{figure*}[t]
		\setlength\tabcolsep{1.5pt} 
	 	\begin{tabular}{ccc}
		\subfigure[SanJose14]{\includegraphics[width = \matrixCellWidth]
			{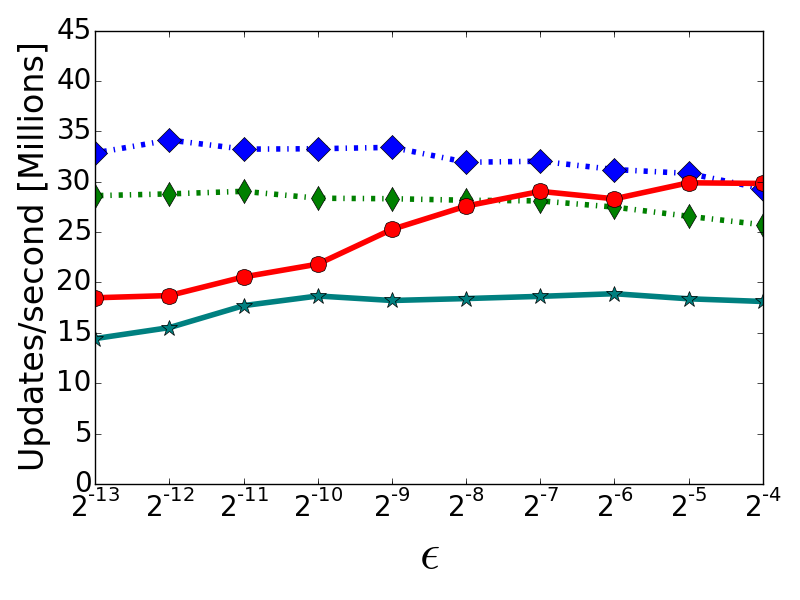}}&
		\subfigure[YouTube]{\includegraphics[width = \matrixCellWidth]
			{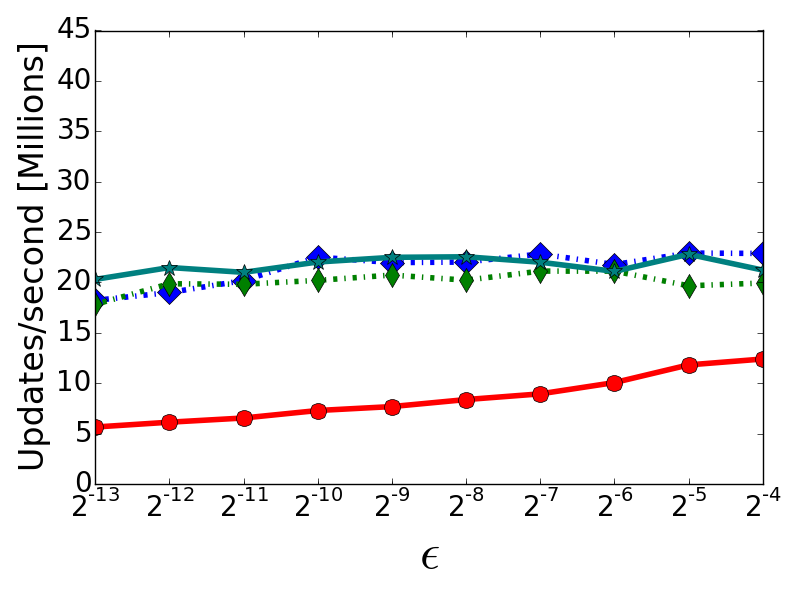}}&
		\subfigure[Chicago16]{\includegraphics[width = \matrixCellWidth]
			{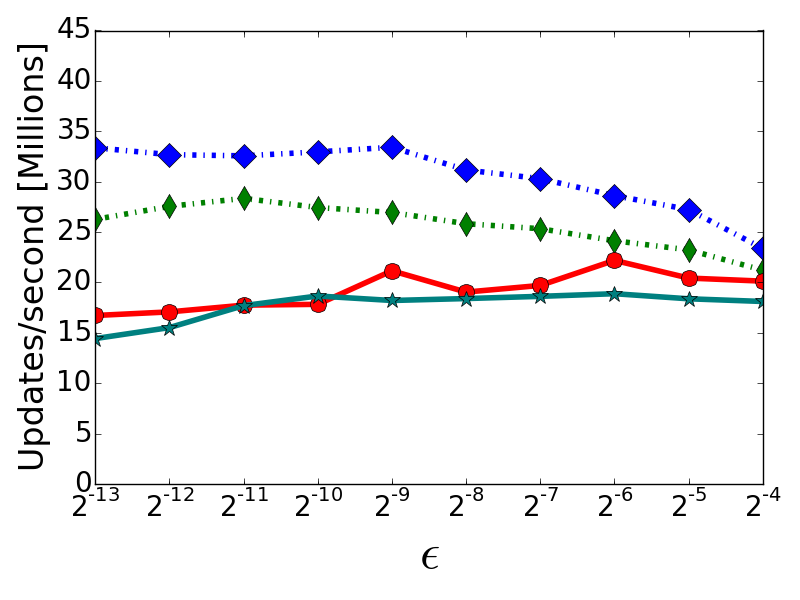}}
		\end{tabular}
		\center{\vspace*{-0.1cm}
		\includegraphics[width =0.80\columnwidth]{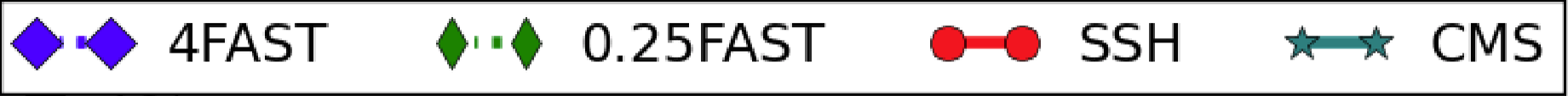}
		\vspace*{-0.1cm}}\\
		\begin{tabular}{ccc}
		\subfigure[SanJose13]{\includegraphics[width = \matrixCellWidth]
			{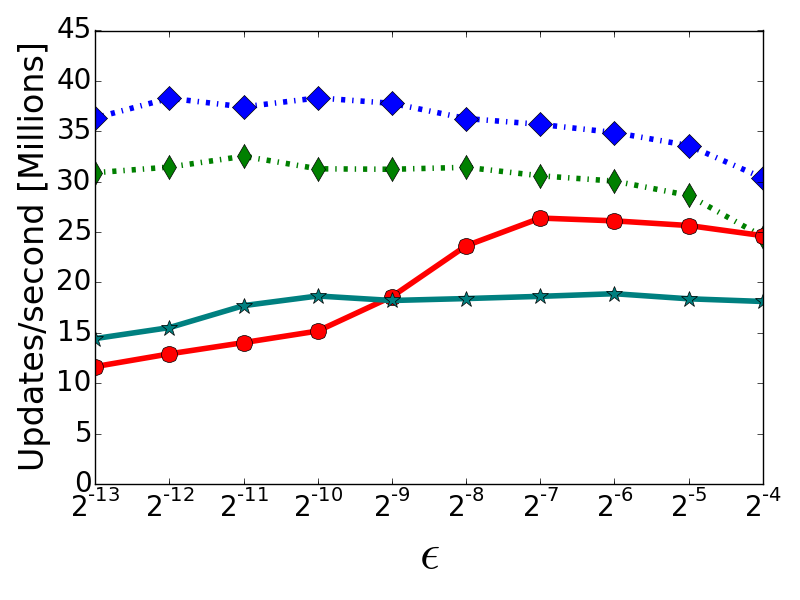}} &
		\subfigure[DC1]{\includegraphics[width =  \matrixCellWidth]
			{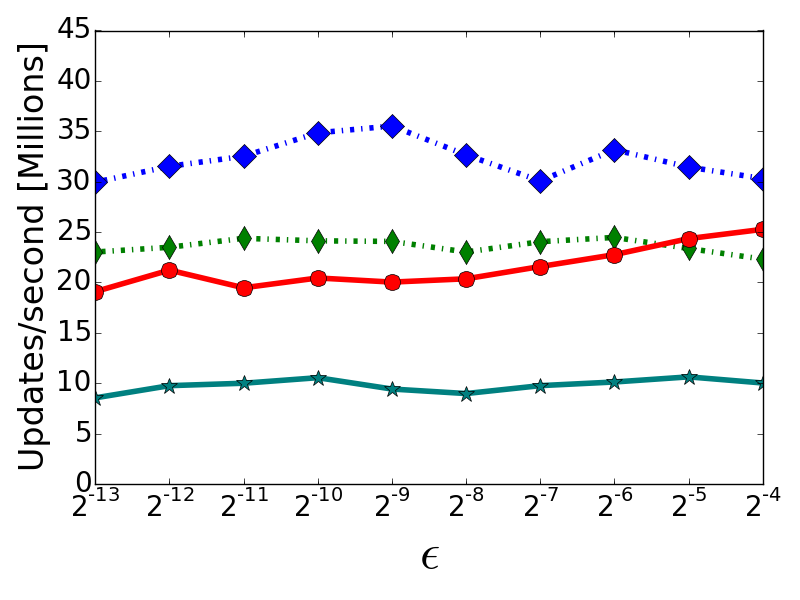}}  &		
		\subfigure[Chicago15]{\includegraphics[width = \matrixCellWidth]
			{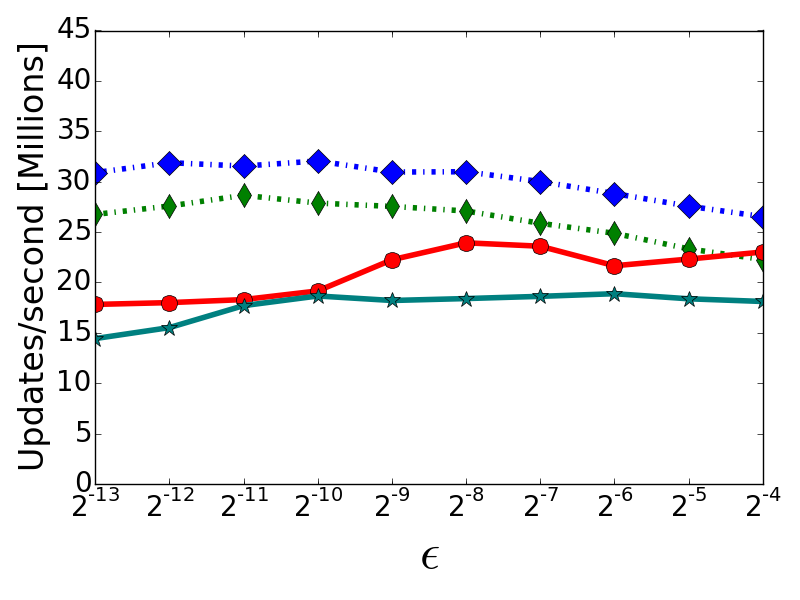}}\\
\ifdefined\EXTENDED
		\subfigure[Zipf0.7]{\includegraphics[width = \matrixCellWidth]
			{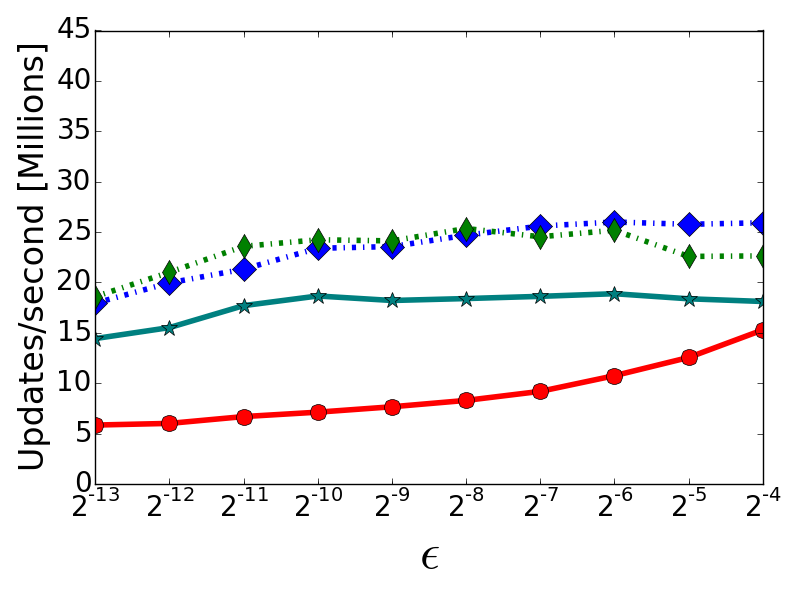}}&
		\subfigure[Zipf1]{\includegraphics[width = \matrixCellWidth]
			{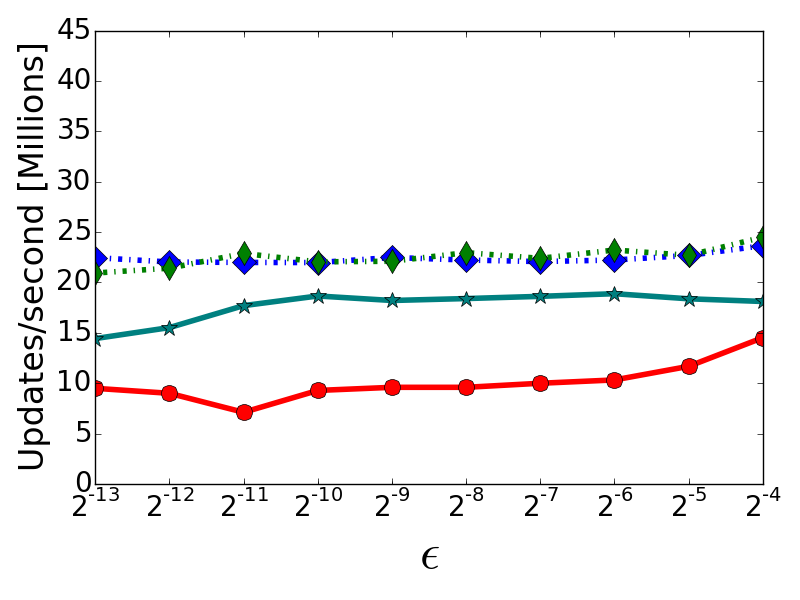}} &
		\subfigure[Zipf1.3]{\includegraphics[width = \matrixCellWidth]
			{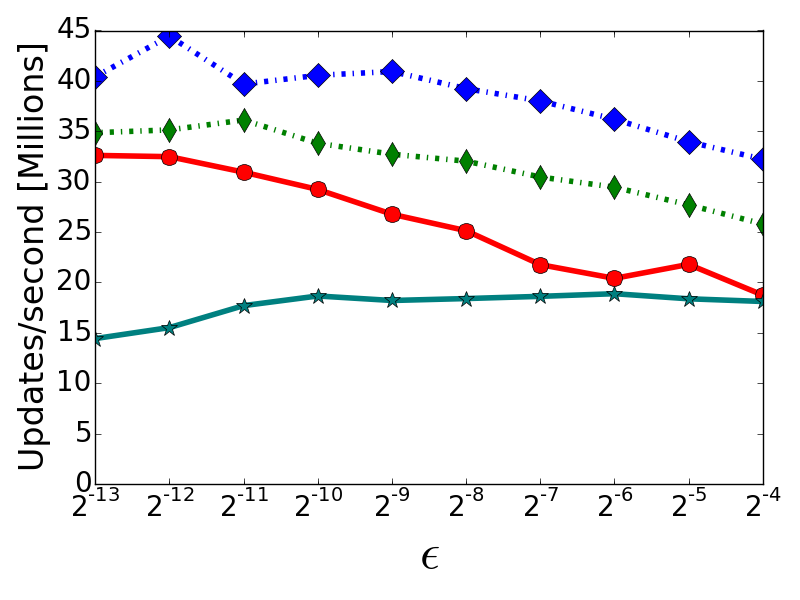}}
\fi
		\end{tabular}
	\caption{Runtime comparison as a function of accuracy guarantee ($\epsilon$) provided by the algorithms.}
	\label{fig:runtime}
\end{figure*}

\section{Hierarchical Heavy Hitters}
\label{sec:HHH}

\emph{Hierarchical heavy hitters (HHH)} algorithms treat IP addresses as a hierarchical domain.
At the bottom are \emph{fully specified} IP addresses such as $p_0 = 101.102.103.104$.
Higher layers include shorter and shorter prefixes of the fully specified addresses.
For example, $p_1 = 101.102.103.*$ and $p_2 = 101.102.*$ are level 1 and level 2 prefixes of $p_0$, respectively.
Such prefixes \emph{generalize} an IP address.
In this example, $p_0 \prec p_1 \prec p_2$, indicating that $p_0$ satisfies the pattern of $p_1$, and any IP address that satisfies $p_1$ also satisfies $p_2$.
The above example refers to a \emph{single dimension} (e.g., the source IP), and can be generalized to \emph{multiple dimensions} (e.g., pairs of source IP and destination IP).
HHH algorithms need to find the heavy hitter prefixes at each level of the induced hierarchy.
For example, this enables identifying heavy hitters subnets, which may be suspected of generating a DDoS attack.
The problem is formally defined in~\cite{HHHMitzenmacher,CormodeHHH}.

\subsection*{Hierarchical Fast (HFAST)}
\emph{Hierarchical FAST (HFAST)} is derived from the algorithm of~\cite{HHHMitzenmacher}.
Specifically, the work of~\cite{HHHMitzenmacher} suggests \emph{Hierarchical Space Saving with a Heap(HSSH)}. 
In their work, the HHH prefixes are distilled from multiple solutions of plain heavy hitter problems.
That is, each prefix pattern has its own separate heavy hitters algorithm that is updated on each packet arrival.
For example, consider a packet whose source IP address is $101.102.103.104$ where the (one dimensional) HHH measurements are carried according to source addresses.
In this case, the packet arrival is translated into the following five heavy hitters update operations:
$101.102.103.104$, $101.102.103.*$, $101.102.*$, $101.*$, and~$*$.
Finally, HHHs are identified by calculating the heavy hitters of each separate heavy hitters algorithm.

H\rss{} is derived by replacing the underlying heavy hitters algorithm in~\cite{HHHMitzenmacher} from Space Saving with heap~\cite{SpaceSavings} to \rss{}.
This asymptotically improves the update complexity from $O\left(H \log\left(\frac{1}{\epsilon}\right)\right)$ to $O\left(H\right)$, where $H$ is the size of the hierarchy.
Since the analysis of~\cite{HHHMitzenmacher} is indifferent to the internal implementation of the heavy hitters algorithm, no  analysis is required for H\rss{}.

Finally, we note that a hierarchical heavy hitters algorithm on sliding windows can be constructed using the work of~\cite{HHHMitzenmacher} by replacing each space saving instance with our W\rss.
The complexity of the proposed algorithm is $O\left(\frac{H}{\epsilon}\right)$ space and $O\left(H\right)$ update time.
To our knowledge, there is no prior work for this problem.

\section{Evaluation}
\label{sec:eval}
\newcommand{\legendWidth}{4.9cm}
\ifdefined \EXTENDED
\newcommand{\windowMatrixCellWidth}{\matrixCellWidth}
\newcommand{\windowMatrixCellHeight}{4.3cm}
\newcommand{\legendHeight}{4.05cm}
\fi
\ifdefined \NINEPAGES
\newcommand{\windowMatrixCellWidth}{5.8cm}
\newcommand{\windowMatrixCellHeight}{3.75cm}
\newcommand{\legendHeight}{3.5cm}
\fi
\begin{figure*}[t]
		\subfigure[SanJose14]{\includegraphics[width = \windowMatrixCellWidth, height = \windowMatrixCellHeight]
			{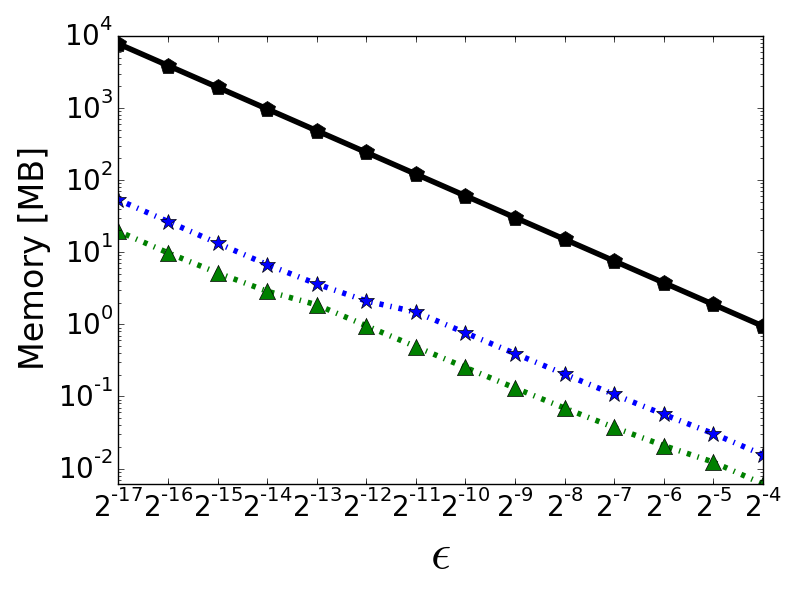}} 	
		\subfigure[YouTube]{\includegraphics[width = \windowMatrixCellWidth, height = \windowMatrixCellHeight]
			{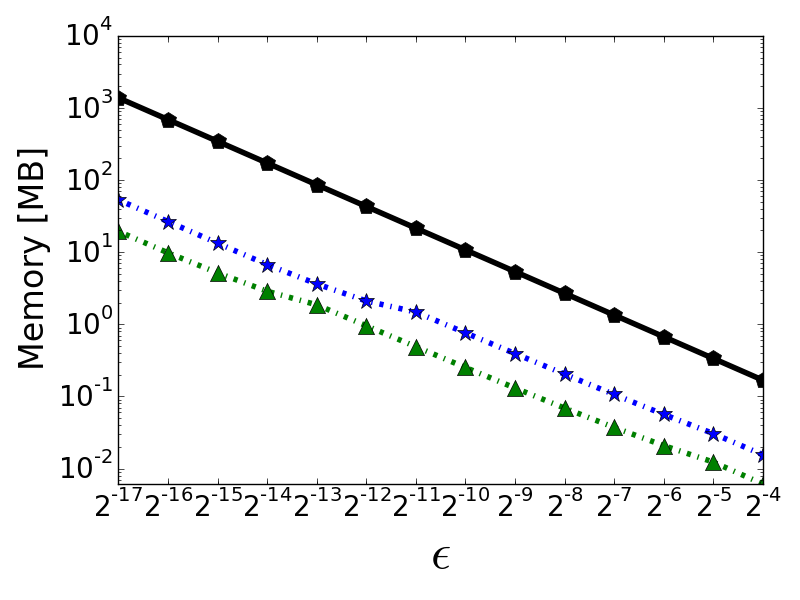}} 	
		\subfigure[Chicago16]{\includegraphics[width = \windowMatrixCellWidth, height = \windowMatrixCellHeight]
			{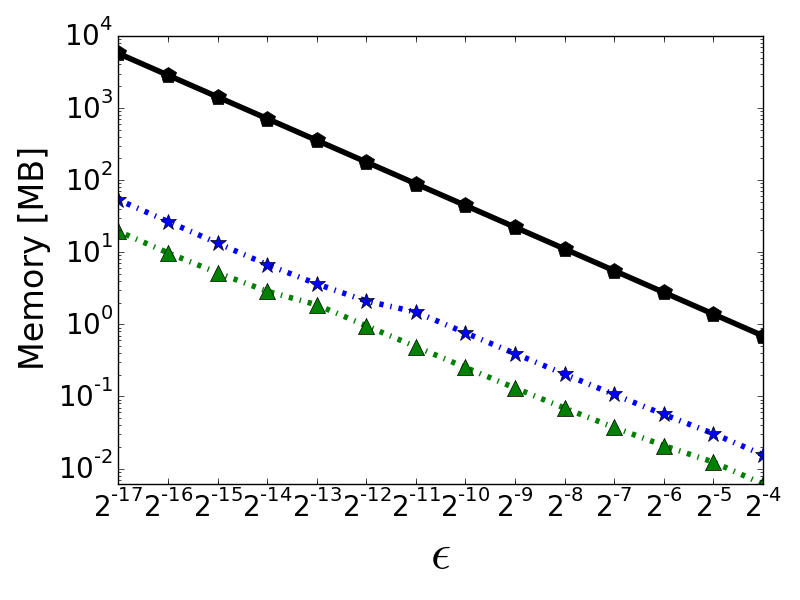}}
		\center{\includegraphics[width=0.8\columnwidth]{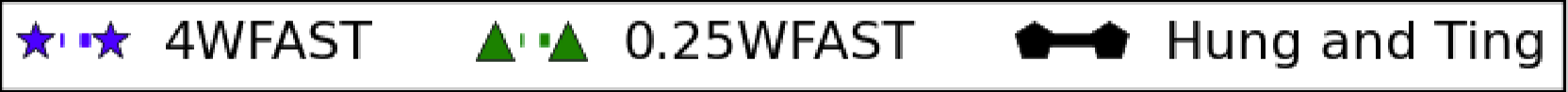}}
	\caption{Space overheads of W\rss{} compared to previous works. Note that W\rss{} operates in constant time while the other algorithm requires linear scanning of all counters.}
    \label{fig:sliwi}
\end{figure*}
\begin{figure*}[t!]
	\captionsetup{justification=centering}
	
	\subfigure[Chicago 16]{\includegraphics[width = \matrixCellWidth]
		{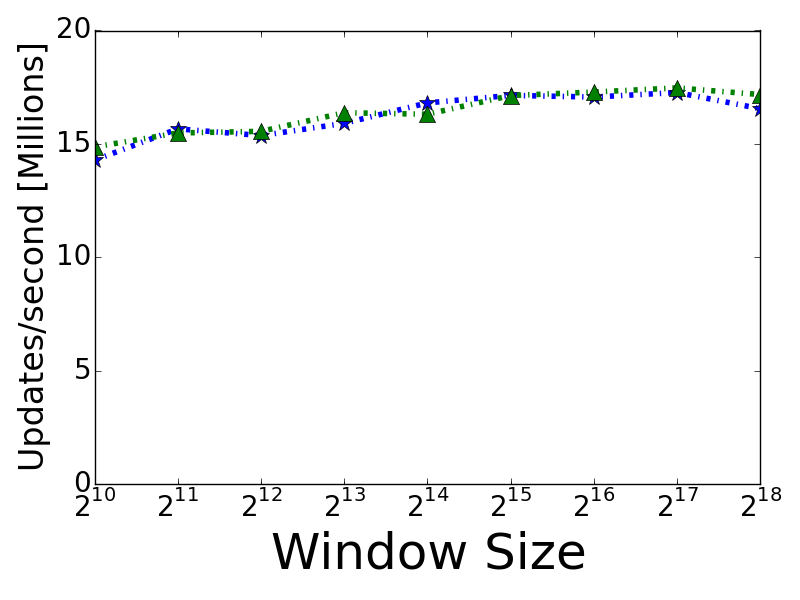}}
	\subfigure[YouTube]{\includegraphics[width = \matrixCellWidth]
		{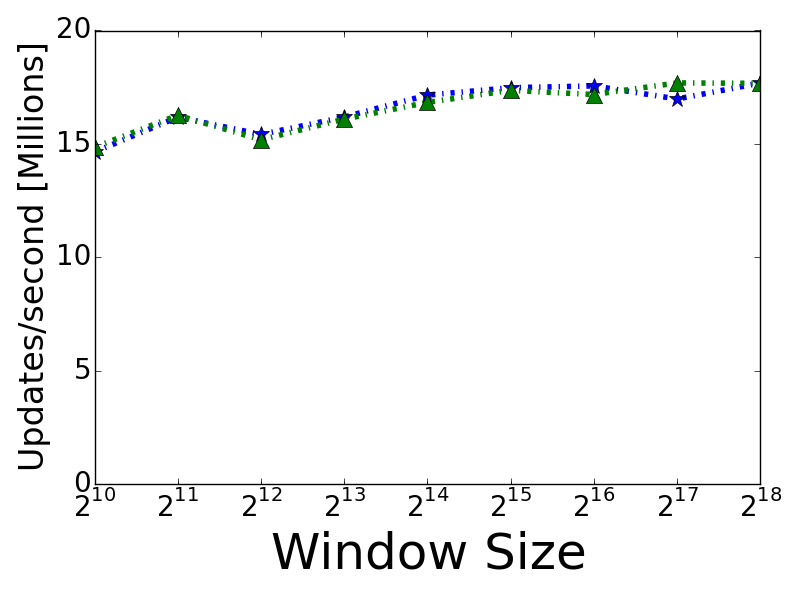}} 	
	\subfigure[DC1]{\includegraphics[width = \matrixCellWidth]
		{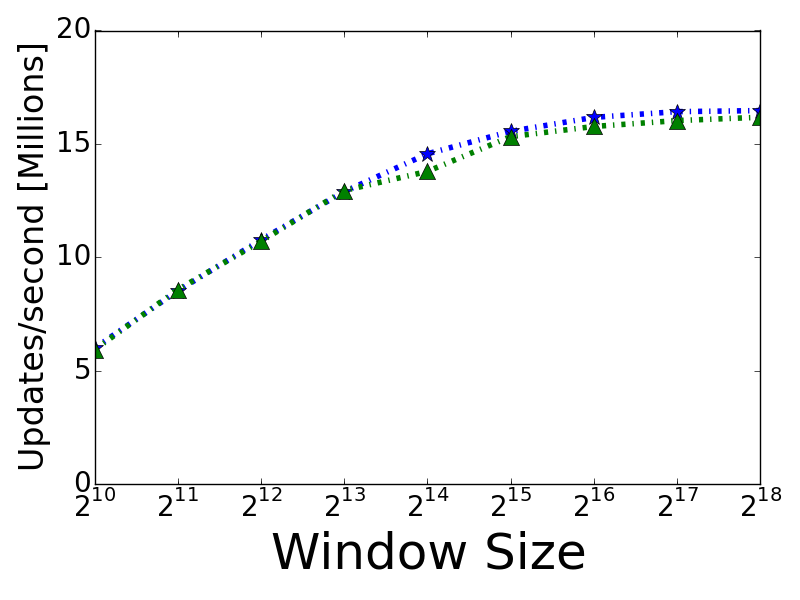}}
	\center{\includegraphics[width=0.8\columnwidth]{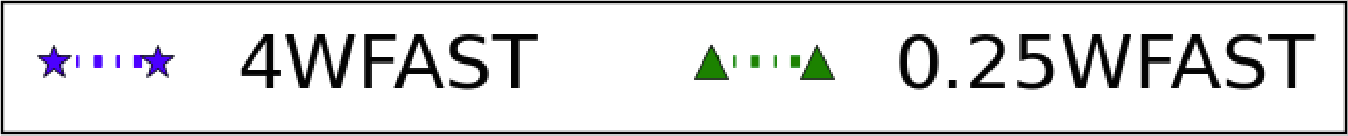}}\\
	\subfigure[Chicago16]{\includegraphics[width = \matrixCellWidth]
		{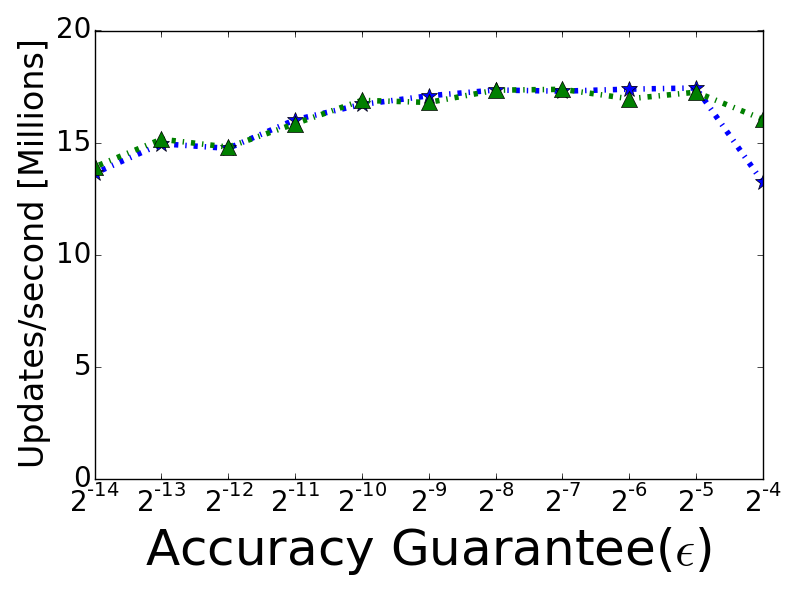}}
	\subfigure[YouTube]{\includegraphics[width = \matrixCellWidth]
		{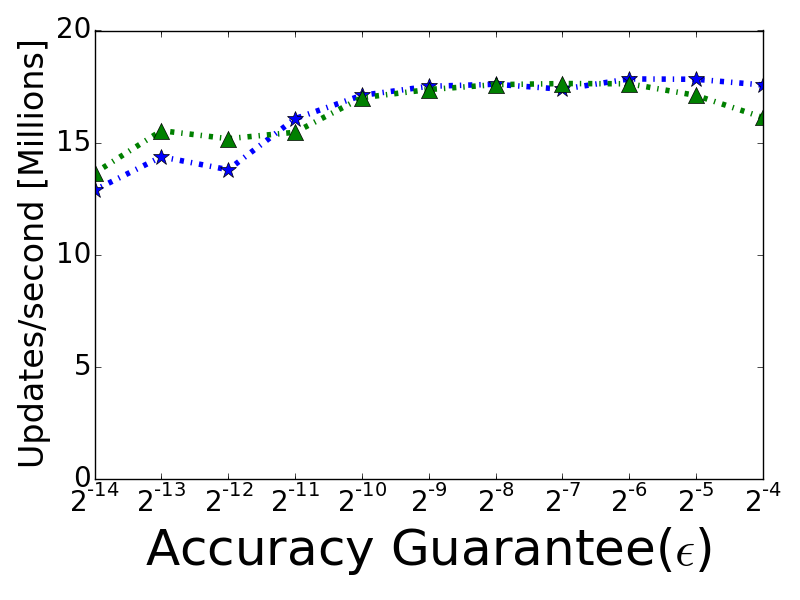}}  		
	\subfigure[DC1]{\includegraphics[width = \matrixCellWidth]
		{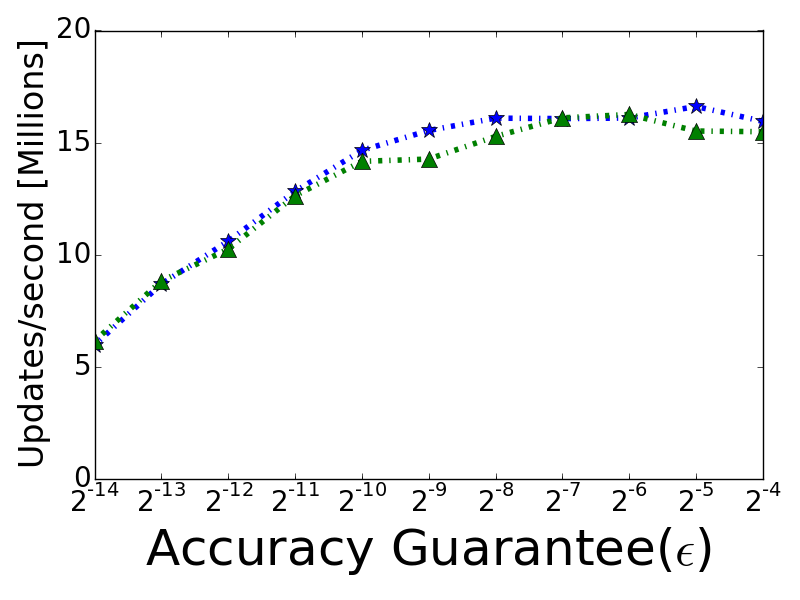}}
	\caption{\label{fig:windowSpeed}
		W\rss{} with varying window sizes ($\varepsilon = 2^{-8}$) and varying $\varepsilon$ (with a window size of $W=2^{16}$). }
\end{figure*}

\begin{figure*}[t]
	\subfigure[SanJose14]{\includegraphics[width = \windowMatrixCellWidth, height = \windowMatrixCellHeight]
		{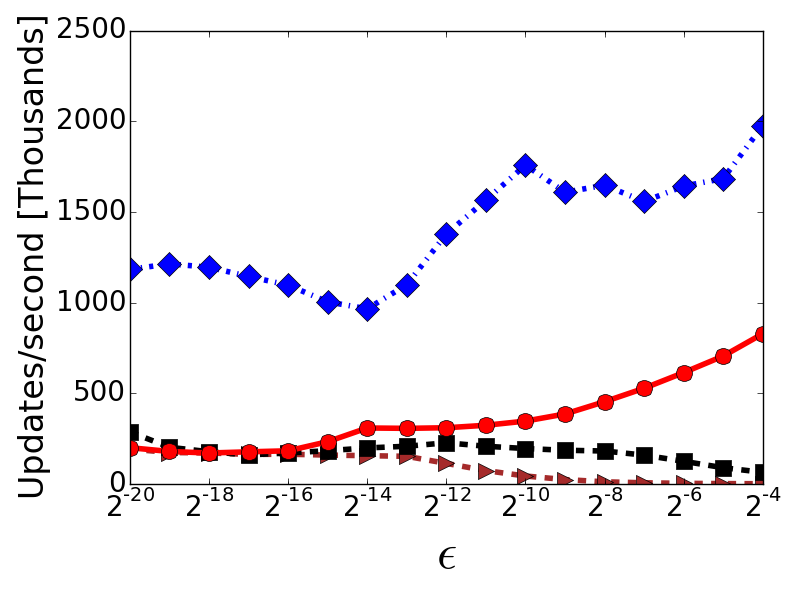}} 	
	\subfigure[Chicago15]{\includegraphics[width = \windowMatrixCellWidth, height = \windowMatrixCellHeight]
		{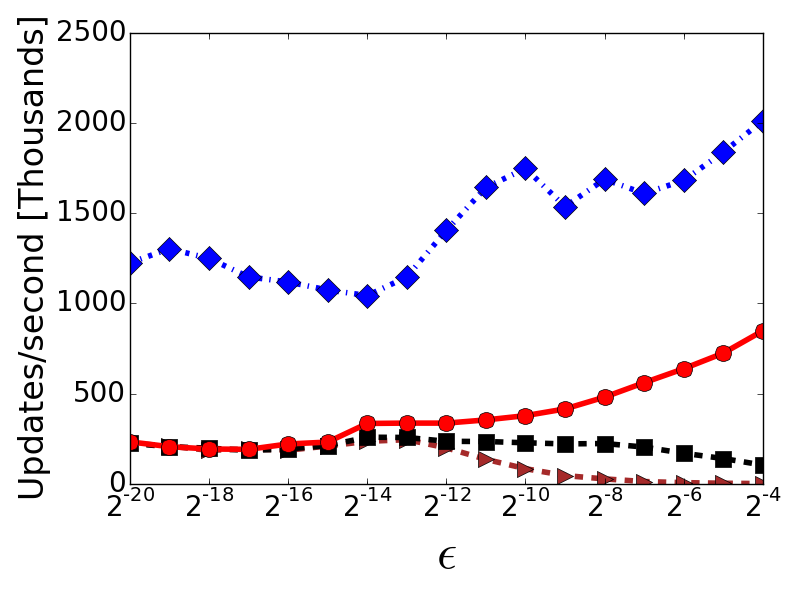}} 	
	\subfigure[Chicago16]{\includegraphics[width = \windowMatrixCellWidth, height = \windowMatrixCellHeight]
		{hhhGraphs/CH16_4-21_hhh.png}}
	\center{\includegraphics[width=1\columnwidth]{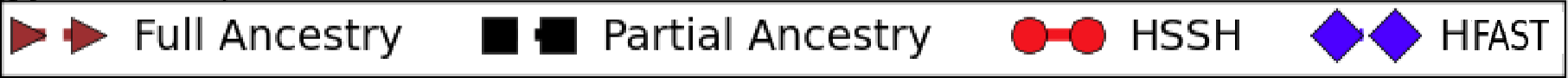}}
	\caption{	\label{fig:HHH} Runtime comparison of HHH algorithms as a function of their accuracy guarantee ($\epsilon$). }
\end{figure*}

Our evaluation is performed on an Intel i7-5500U CPU with a clock speed of 2.4GHz, 16 GB RAM and a Windows 8.1 operating system.
We compare our C++ prototypes to the following alternatives:

\emph{Count Min Sketch (CMS)}~\cite{CMSketch} -- a sketch based solution that can only solve the volume estimation problem.

\emph{Space Saving Heap (SSH)} -- a heap based implementation~\cite{SpaceSavingIsTheBest2010} of Space Saving~\cite{SpaceSavings} that has a logarithmic runtime complexity.

\emph{Hierarchical Space Saving Heap (HSSH)} -- a hierarchical heavy hitters algorithm~\cite{HHHMitzenmacher} that uses SSH as a building block and operates in $O(H\log(\frac{1}{\varepsilon}))$ complexity.

\emph{Full Ancestry} -- a trie based HHH algorithm suggested by~\cite{CormodeHHH}, which operates in $O\left(H \log{\epsilon N}  \right)$ complexity.

\emph{Partial Ancestry} -- a trie based HHH algorithm suggested by~\cite{CormodeHHH}, which operates in $O\left(H \log{\epsilon N}  \right)$ complexity and is considered faster than Full Ancestry.

Related work implementations were taken from open source libraries released by~\cite{SpaceSavingIsTheBest} for streams and by~\cite{HHHMitzenmacher} for hierarchical heavy hitters.
As we have no access to a concrete implementation of a competing sliding window protocol, we compare W\rss{} to Hung and Ting's algorithm~\cite{HungAndTing} by conservatively estimating the space needed by their approach.
Each data point we report here is the average of 10 runs.


\subsection{Datasets}
\ifdefined \NINEPAGES

\fi
 \begin{table*}[]
 \scriptsize
 	\addtolength{\tabcolsep}{-0pt} \addtolength{\parskip}{-0.2mm} \center{
 		\begin{tabular}{|c|c|c|c|c|c|c|c|}
 			\hline
 			Trace &Date(Y/M/D) &  \#Packets & Total volume & Mean size & Max size & \% large packets  & \% large packet traffic\tabularnewline
 			\hline
 			\hline
 			Chicago16 & 2016/02/18 & 97 M & 94 GB & 1046 & 49458 & 0.34\% & 0.5\%\tabularnewline
 			\hline
 			Chicago15 & 2015/12/17 & 85 M & 80 GB & 1013 & 64134 & 0.22\%  & 0.34\%\tabularnewline
 			\hline
 			SanJose14 & 2014/06/19 & 112 M & 149 GB & 1424 & 65535 & 0.78\% & 25.02\%\tabularnewline
 			\hline
 			SanJose13 & 2013/12/19 & 97 M & 110 GB & 1225 & 65528 & 0.49\% & 18.81\% \tabularnewline
 			\hline
 			DC1 & 2010 & 7.3M & 6.1GB & 894 & 1476 & 0\% & 0\%
			\tabularnewline
			\hline 			
 		\end{tabular}
 		\normalsize
 		{{\caption{A summary of key characteristics of the real Internet traces used in this work. }
 				\label{tbl:traces}
 			}}}
 		\end{table*}
\normalsize

Our evaluation includes the following datasets. The packet traces characteristics are summarized in Table~\ref{tbl:traces}.


The CAIDA  backbone Internet traces that monitor links in Chicago~\cite{CAIDACH15,CAIDACH16} and San Jose~\cite{CAIDASJ13,CAIDASJ14}.
\ifdefined\EXTENDED
The data includes a mix of UDP/TCP and ICMP packets.
The Chicago16~\cite{CAIDACH16} and Chicago15~\cite{CAIDACH15} data sets were taken from the `equinix-chicago' high-speed monitor. 
Similarly, the SanJose14~\cite{CAIDASJ14} and SanJose13~\cite{CAIDASJ13} were taken from the `equinix-sanjose' monitor.
\fi
A datacenter trace from a large university~\cite{Benson2010DC} and a trace of 436K YouTube video accesses~\cite{YouTubeDataset}. The weight of a video is its length in~seconds.
\ifdefined\EXTENDED
\item
Self generated synthetic traces following a Zipfian distribution with varying skews.
A trace of skew $X$ is denoted \emph{Zipf$X$}.
Each trace is unweighed (each element has weight $1$) and contains 10M elements.
\fi

As shown in Table~\ref{tbl:traces}, the impact of jumbo frames varies between backbone links.
Yet, the weight of large packets increases over time in both.
In the San Jose link, the number and volume of large packets have increased by 50\% within a period of 6 months.
In the Chicago link, large packets are still insignificant, but their number and volume have increased by 50\% in two months.

\subsection{Effect of $\gamma$ on Runtime}
\ifdefined \EXTENDED
We begin the evaluation by exploring our trade-off parameter $\gamma$.
Recall that smaller $\gamma$ yields space efficiency while the runtime is proportional to $\frac{1}{\gamma}$, i.e, smaller $\gamma$ is expected to cause a slower runtime.
Figure~\ref{fig:gamma} shows runtime performance of \rss{} as a function of $\gamma$ for three different $\varepsilon$ values ($2^{-8},2^{-10},2^{-12}$).
As can be observed, in practice, we indeed get speedup with larger $\gamma$ values.
But, we reach a saturation point and increasing $\gamma$ beyond a certain threshold has little impact on performance.
It is encouraging that even with small values of $\gamma$ such as $2^{-7}$, \rss{} is still reasonably fast.
\fi
\ifdefined \NINEPAGES
Recall that smaller $\gamma$ yields space efficiency while the runtime is proportional to $\frac{1}{\gamma}$, i.e, smaller $\gamma$ is expected to cause a slower runtime.
In Appendix~\ref{sec:missing-figure}, we show runtime performance evaluation of \rss{} as a function of $\gamma$ for three different $\varepsilon$ values ($2^{-8},2^{-10},2^{-12}$).
While we indeed obtained a speedup with larger $\gamma$ values, increasing $\gamma$ beyond a certain small threshold has little impact on performance.
\fi
For the rest of our evaluation, we focus on $\gamma = 0.25$ that offers attractive space/time trade off, as well as on $\gamma = 4$ that yields higher performance at the expense of more space.

\ifdefined \EMPIRICAL
	\subsection{Effect of $\gamma$ on empirical error}
	\ifdefined \NINEPAGES
	For comparing the empirical error, we fixed the maximal allowed error to be $\epsilon=2^{-8}$.
    We configured \rss{} with various values of $\gamma$ and compared it to SSH with similar number of counters (i.e., with $\epsilon_{SSH} = 2^{-8}/(1+\gamma)$).
	The experiment results, presented in detail in the full version of the paper~\cite{fullVersion}, indicate that for a given number of counters, \rss{} and SSH have roughly the same estimation quality.
	\fi
	\ifdefined \EXTENDED
	We continue by comparing our empirical error to that of Space Saving, which is considered the state of the art.
	Since \rss{} requires more counters than Space Saving for the same $\epsilon$, we configured Space Saving with a smaller $\epsilon$ so that the total number of counters is equal to that of \rss.
    In Figure~\ref{fig:emperror}, \rss{} is configured with $\epsilon = 2^{-8}$ and Space Saving is configured with $\epsilon=\frac{1}{256\cdot{(1+\gamma)}}$.
	
	The error is calculated according to the \emph{On Arrival} model suggested by~\cite{WCSS}.
	In this model, we perform a query prior to each packet arrival and calculate the \emph{root mean square error} of these queries.
	The error is averaged after 10 arbitrary intervals of contiguous 2M packets.

	As can be observed in Figure~\ref{fig:emperror}, \rss{} is indeed comparable to Space Saving in terms of empirical error.
	Interestingly, in some cases \rss{} is slightly better than Space Saving.
	We believe that this is due to the \emph{eviction policy}.
	Specifically, removing the counter with minimal value does not always yield the best empirical error.
	In our implementation, items of the same (rounded) volume are evicted according to LIFO (last in first out) for runtime reasons.
	We leave it for future work to optimize the eviction policy of \rss.
	\fi
\fi

\subsection{Speed vs. Space Tradeoff}
To explain the tradeoff proposed by \rss, we measured the runtime of the various algorithms for a fixed error guarantee.
Here, SSH and CMS are fully determined by the error guarantees (set to be $\epsilon=2^{-8}$) and thus have a single measurement point.
CMS requires more counters as it uses $10$ rows of $\lceil e/\epsilon\rceil$ counters each, while SSH only requires $1/\epsilon$.
\rss{} can provide the same error guarantee for different $\gamma$ values, which affects both runtime and the number of counters.
Hence, \rss{} is represented by a curve.
As Figure~\ref{fig:tradeoff} shows, in all traces, allocating a few additional counters to the $1/\epsilon$ required by SSH allows \rss{} to achieve higher throughput.
Additionally, on all traces, \rss{} provides faster throughput than CMS with far fewer counters.
While \rss{} has larger per counter overheads than CMS, its ID to counter mapping allows it to solve the {\sc {Weighted Heavy Hitters}} problem that CMS cannot.


\subsection{Operation Speed Comparison}
Figure~\ref{fig:runtime} presents a comparative analysis of the operation speed of previous approaches.
Recall that CMS is a probabilistic scheme; we configured it with a failure probability of $0.1\%$.
For \rss, we used two configurations: $\gamma=4$ (4\rss) and $\gamma=0.25$~(0.25\rss).

As can be observed, 4\rss{} and 0.25\rss{} are considerably faster than the alternatives in Chicago16 and YouTube.
In SanJose14 and SanJose13, SSH is as fast as 4\rss{} for a large $\epsilon$ (small number of counters).
Yet, as $\epsilon$ decreases and the number of counters increases, SSH becomes slower due to its logarithmic complexity.
In contrast, CMS is almost workload independent.
When considering only previous work, in some workloads CMS is faster than SSH, mainly because SSH's performance is workload dependent.
\ifdefined\EXTENDED
The bottom 3 figures (g,h,i) show results for synthetic unweighted Zipf traces with skew parameters of $0.7,1,1.3$, respectively.
As can be observed, for mildly skewed distributions, CMS is faster than SSH, while for skewed distributions such as when the skew is 1.3, SSH is faster.
In all these measurements, 4\rss{} is faster than the alternatives.
\fi

\subsection{Sliding Window}
We evaluate W\rss{} compared to Hung and Ting's algorithm~\cite{HungAndTing}, which is the only one that supports weighted updates on sliding windows.
Figure~\ref{fig:sliwi} shows the memory consumption of W\rss{} with parameters $\gamma =4$~and~$\gamma =0.25$ (4W\rss, 0.25\rss) compared to Hung and Ting's algorithm.  All algorithms are configured to provide the same worst case error guarantee.
As shown, W\rss{} is up to 100 times more space efficient than Hung and Ting's algorithm.
Sadly, we could not obtain an implementation of Hung and Ting's algorithm and thus do not compare its runtime to W\rss{}.
However, W\rss{} improves their update complexity from $O(\frac{A}{\epsilon})$, where $A$ is the average packet size, to $O(1)$.

Figure~\ref{fig:windowSpeed} shows the operation speed of W\rss{} for different window sizes and different $\varepsilon$ values.
\ifdefined\EXTENDED
As seen, W\rss{} achieves over 15 million updates per second using a single thread.
It is about half as fast as \rss{} for streams and still within the range of acceptable parameters.
\fi
There is little dependence in window size and $\varepsilon$ with the exception of the DC1 dataset.
In this dataset, since the average and maximal packet sizes are similar, the inner working of W\rss{} causes overflows to be more frequent when $\varepsilon$ is close to the window size.
Thus, to achieve similar performance as the other traces one needs sufficiently large window size in this trace.

\subsection{Hierarchical Heavy Hitters}

In Figure~\ref{fig:HHH}, we evaluate the speed of our H\rss{} compared to the algorithm of~\cite{HHHMitzenmacher}, which is denoted by HSSH, as well as the Partial Ancestry and Full Ancestry algorithms by~\cite{CormodeHHH}.
We used the library of~\cite{HHHMitzenmacher} for their own HSSH implementation as well as for the Partial Ancestry and Full Ancestry implementations.
Since the library was released for Linux, we used a different machine for our H\rss{} evaluation.
Specifically, we used a Dell 730 server running Ubuntu 16.04.01 release.
The server has 128GB of RAM and an Intel(R) Xeon(R) CPU E5-2667 v4 @ 3.20GHz processor.

We used two dimensional source/destination hierarchies in byte granularity, where networks IDs are assumed to be 8, 16 or 24 bits long.
The weight of each packet is its byte volume, including both the payload size and the header size.
As depicted, H\rss{} is up to 7 times faster than the best alternative and at least 2.4 times faster in every data point.
It appears that for large $\epsilon$ values, HSSH is faster than the Partial and Full Ancestry algorithms.
Yet, for small $\epsilon$ values, all previous algorithms operate in similar speed. 

\section{Discussion}
\label{sec:discussion}

In this paper, we presented algorithms for estimating per flow traffic volume in streams, sliding windows and hierarchical domains.
Our algorithms offer both asymptotic and empirical improvements for these problems.

For streams, \rss{} processes packets in constant time while being asymptotically space optimal. 
This is enabled by our novel approach of maintaining only a partial order between counters.
An evaluation over real-world traffic traces 
\ifdefined\EXTENDED
as well as synthetic ones,
\fi 
has yielded a speed improvement of up to 2.4X compared to previous work. 
\ifdefined\EXTENDED
This is significant since the combination of fast line rates with NFV trends imposes strict timing constraints.
\fi

In the sliding window case, we showed that W\rss{} works reasonably fast and offers 100x reduction in required space, bringing sliding windows to the realm of possibility.
For a given error of $W\cdot M\cdot \eps$, W\rss{} requires $O\parentheses{\frac{1}{\epsilon}}$ counters while previous work uses $O\parentheses{\frac{A}{\epsilon}}$, where $A$ is the average packet size.
Moreover, W\rss{} runs in constant time while previous work runs in $O\parentheses{\frac{A}{\epsilon}}$.

For hierarchical domains, we presented H\rss{} that requires $O(\frac{H}{\epsilon})$ space and has $O(H)$ update complexity.
This improves over the $O\left(H \log\frac{1}{\epsilon}\right)$ update complexity of previous work.
Additionally, we demonstrated a speedup of 2.4X-7X on real Internet traces.
To our knowledge, there is no prior work on that problem and we plan to examine its possible applications in the future. The code of FAST is available as open source~\cite{FASTCode}. 

\thanks{We thank Yechiel Kimchi for helpful code optimization~suggestions.}




{ 
	\bibliographystyle{acm}
\scriptsize
	\bibliography{refs}
}

\newpage
\appendix

\begin{figure*}[t]
	
	\subfigure[SanJose14]{\includegraphics[width = \matrixCellWidth]
		{gammaGraphs/SJ14_newGraphs_gammaRun_gamma.png}}	
	\subfigure[YouTube]{\includegraphics[width = \matrixCellWidth]
		{gammaGraphs/YouTube_newGraphs_gammaRun_gamma.png}}	
	\subfigure[Chicago16]{\includegraphics[width = \matrixCellWidth]
		{gammaGraphs/CH16_newGraphs_gammaRun_gamma.png}}
	\center{\includegraphics[width=0.8\columnwidth]{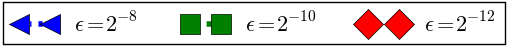}}\\
	\subfigure[SanJose13]{\includegraphics[width = \matrixCellWidth]
		{gammaGraphs/SJ13_newGraphs_gammaRun_gamma.png}}
	\subfigure[DC1]{\includegraphics[width =  \matrixCellWidth]
		{gammaGraphs/Univ1_newGraphs_gammaRun_gamma.png}} 		
	\subfigure[Chicago15]{\includegraphics[width = \matrixCellWidth]
		{gammaGraphs/CH15_newGraphs_gammaRun_gamma.png}}
	\caption{	\label{fig:gamma}The effect of parameter $\gamma$ on operation speed for different error guarantees ($\epsilon$). $\gamma$ influences the space requirement as the algorithm is allocated with $\nrCounters$ counters.}
\end{figure*}
\section{Missing Proofs}
\label{sec:missing-proofs}
\subsection*{Proof of Lemma~\ref{lem:lowerBound}\\}
\begin{proof}
We prove $v_{x,t}\le q_t(x)$ by induction over $t$. \\
\textbf{Basis:} $t = 0$. Here, we have $v_{x,t} = 0 = q_t(x)$. \\
\textbf{Hypothesis:} $v_{x,t-1}\le q_{t-1}(x)$\\
\textbf{Step:} $\angle{x_t,w_t}$ arrives at time $t$. By case~analysis:

Consider the case where the queried item $x$ is not the arriving one (i.e., $x\neq x_t$).
In this case, we have $v_{x,t} = v_{x,t-1}$.
If $x\in C_{t-1}$ but was evicted (Line~\ref{line:eviction}) then $c_x\in \text{argmin}_{y\in C_{t-1}}(c_{y,t-1})$. This means that:
\begin{multline*}
q_{t-1}(x)= r_{x,t-1}+\stepLetter\cdot\text{argmin}_{y\in C_{t-1}}(c_{y,t-1})\\
\le \stepLetter-1+\stepLetter\cdot\text{argmin}_{y\in C_t}(c_{y,t}) = q_t(x),
\end{multline*}
\normalsize
where the last equation follows from the query for $x\notin C_t$ (Line~\ref{line:noCounterQuery}).
Next, if $x\in C_{t-1}$ and $x\in C_t$, its estimated volume is determined by Line~\ref{line:normalQuery} and we get $q_t(x)=q_{t-1}(x) \ge v_{x,t-1}=v_{x,t}$.
If $x\notin C_{t-1}$ then $x\notin C_{t}$, so the values of $q_t(x),q_{t-1}(x)$ are determined by line~\ref{line:noCounterQuery}.
Since the value of $\min_{y\in C} c_y$ can only increase over time, we have $q_t(x)\ge q_{t-1}(x)\ge v_{x,t}$ and the claim holds.

On the other hand, assume that we are queried about the last item, i.e., $x=x_t$.
In this case, we get $v_{x,t} = v_{x,t-1}+w_t$.
We consider the following cases:
First, if $x\in C_{t-1}$, then $q_t(x) = q_{t-1}(x) + w_t$.
Using the hypothesis, we conclude that $v_{x,t}=v_{x,t-1}+w_t\le q_{t-1}(x)+w_t = q_t(x)$ as required.
Next, if $|C_{t-1}| < \nrCountersLetter$, we also have $q_t(x) = q_{t-1}(x) + w_t$ and the above analysis holds.
Finally, if $x\notin C_{t-1}$ and $|C_{t-1}|=\nrCountersLetter$, then
\footnotesize
\begin{align}
q_{t-1}(x) = s - 1 + s\cdot \min_{y\in C_{t-1}}c_{y,t-1}.\label{eq:qt1query}
\end{align}
\normalsize
On the other hand, when $x$ arrives, the condition of Line~\ref{line:firstIf} was not satisfied, and thus
\footnotesize
\begin{align*}
q_t(x) &= r_{x,t}+s\cdot c_{x,t} =(\stepLetter-1+w)\mod \stepLetter \\&\qquad+ s\cdotpa{ \min_{y\in C_{t-1}}c_{y,t-1} + \floor{	\frac{\stepLetter - 1 + w}{\stepLetter}}}\\
_{\parentheses{\tiny \text{Observation~\ref{obs:modulo}}}}&= s\cdot \min_{y\in C_{t-1}}c_{y,t-1} + \stepLetter - 1 + w\\
_{\eqref{eq:qt1query}}&=q_{t-1}(x) + w\\
_{\parentheses{\substack{\tiny \text{induction}\\\text{hypothesis}}}}&\ge v_{x,t-1} + w = v_{x,t}.\qquad\qquad\qquad\qquad\qed
\end{align*}
\normalsize
\end{proof}

\subsection*{Proof of Lemma~\ref{lem:exactCount}\\}
\begin{proof}
	Since $|C|\le\nrCountersLetter$, we get that the conditions in Line~\ref{line:firstIf} and Line~\ref{line:normalQuery} are always satisfied. Before the queried element $x$ first appeared, we have $r_x=c_x=0$ and thus \query$=0$. Once $x$ appears once, it gets a counter and upon every arrival with value $w$, the estimation for $x$ exactly increases by $w$, since $x$ never gets evicted (which can only happen in Line~\ref{line:min}).
\end{proof}
\newpage
\subsection*{Proof of Lemma~\ref{lem:sum-of-counters}\\}
\begin{proof}
	We prove the claim by induction on the stream length~$t$.
 \textbf{Basis:} $t = 0$.\\
		In this case, all counters have value of $0$ and thus  \\$\sum_{x\in C_t} q_t(x) = 0 = t\cdotpa{ M\cdot (1+\gamma/2)}$.\\
\textbf{Hypothesis:} \footnotesize$\sum_{x\in C_{t-1}} q_{t-1}(x) \le (t-1)\cdot{ M\cdot (1+\gamma/2)}$\normalsize.\\
\textbf{Step:} $\angle{x_t,w_t}$ arrives at time $t$.
		We consider the following cases:
		\begin{enumerate}
			\item $x\in C_{t-1}$ or $|C_{t-1}|<\nrCounters$. In this case, the condition in Line~\ref{line:firstIf} is satisfied and thus $c_{x,t} = c_{x,t-1} + \floor{	\frac{r_{x,t-1} + w}{\stepLetter}}$ (Line~\ref{line:regularCounterUpdate}) and $r_{x,t} = (r_{x,t-1} + {w}) \mod {\stepLetter}$ (Line~\ref{line:regularRemainderUpdate}). By Observation~\ref{obs:modulo} we get
			\scriptsize
			\begin{align}
			\hspace{-0.0cm} q_t(x) &=_{\parentheses{\substack{\text{by line}\\\ref{line:normalQuery}}}} r_{x,t} + \stepLetter\cdot c_{x,t} \notag\\
			&= c_{x,t-1} + \floor{	\frac{r_{x,t-1} + w}{\stepLetter}} + (r_{x,t-1} + {w}) \mod {\stepLetter}\notag\\
			&\hspace{-0.0cm}= w + c_{x,t-1} + r_{x,t-1} = q_{t-1}(x) + w\label{eq:qt-val}.
			\end{align}
			\normalsize
			Since the value of a query for every $y\in C_t\setminus\{x\}$ remains unchanged, we get that
			\scriptsize
			\begin{align*}
			\sum_{y\in C_t} q_t(y) &= q_t(x) + \sum_{\substack{y\in C_{t-1}\\ y\neq x}} q_{t-1}(y) \\
			_{(\text{by }\eqref{eq:qt-val})} &= w + q_{t-1}(x) + \sum_{\substack{y\in C_{t-1}\\ y\neq x}} q_{t-1}(y)\\
			&= w + \sum_{\substack{y\in C_{t-1}}} q_{t-1}(y)\\
			_{\parentheses{\substack{\tiny \text{induction}\\\text{hypothesis}}}}&\le w + (t-1)\cdotpa{ M\cdot (1+\gamma/2)}\\
			&\le M + (t-1)\cdotpa{ M\cdot (1+\gamma/2)}\\
			_{\parentheses{\gamma\ge 0}}&\le t\cdotpa{ M\cdot (1+\gamma/2)}.
			\end{align*}
			\normalsize
			\item $x\notin C_{t-1}$ and $|C_{t-1}|=\nrCounters$. In this case, the condition of Line~\ref{line:firstIf} is false and therefore $c_{x,t} = c_{m,t-1} + \floor{	\frac{\stepLetter - 1 + w}{\stepLetter}}$~(Line~\ref{line:takeoverCounterUpdate}) and $r_{x,t} \gets (\stepLetter - 1 + w) \mod \stepLetter$~(Line~\ref{line:takeoverRemainderUpdate}).
			From Observation~\ref{obs:modulo} we get that
			\footnotesize
			\begin{align}
			\hspace{-0.0cm} q_t(x) &=_{\parentheses{\substack{\text{by Line}\\\ref{line:normalQuery}}}} r_{x,t} + \stepLetter\cdot c_{x,t} \notag\\
			&= c_{m,t-1} + \floor{	\frac{\stepLetter - 1 + w}{\stepLetter}} + (\stepLetter - 1 + {w}) \mod {\stepLetter}\notag\\
			&\hspace{-0.0cm}= w + c_{m,t-1} + \stepLetter - 1\notag\\
			& = q_{t-1}(m) - r_{m,t-1} + \floor{\frac{M\gamma}{2}} + w\notag\\
			&\le q_{t-1}(m) + \floor{\frac{M\gamma}{2}} + w\label{eq:qt-val}.
			\end{align}
			\normalsize
			As before, the value of a query for every $y\in C_t\setminus\{x\}$ is unchanged, and since $C_{t-1}\setminus C_t = \{m\}$,
			\scriptsize
			\begin{align*}
			\sum_{y\in C_t} q_t(y) &= q_t(x) - q_{t-1}(m) + \sum_{\substack{y\in C_{t-1}}} q_{t-1}(y)  \\
			_{(\text{by }\eqref{eq:qt-val})} &\le \floor{\frac{M\gamma}{2}} + w + \sum_{\substack{y\in C_{t-1}}} q_{t-1}(y) \\
			_{\parentheses{\substack{\tiny \text{induction}\\\text{hypothesis}}}}&\le \floor{\frac{M\gamma}{2}} + w + (t-1)\cdotpa{ M\cdot (1+\gamma/2)}\\
			&\le \floor{\frac{M\gamma}{2}} + M + (t-1)\cdotpa{ M\cdot (1+\gamma/2)}\\
			_{\parentheses{\gamma\ge 0}}&\le t\cdotpa{ M\cdot (1+\gamma/2)}.\qquad\qquad\qquad\qquad\qed
			\end{align*}
			\normalsize
		\end{enumerate}
\end{proof}

\subsection*{Proof of Lemma~\ref{lem:upperBound}\\}
\begin{proof}
First, consider the case where the stream contains at most $\nrCounters$ distinct elements.
By Lemma~\ref{lem:exactCount}, $\widehat{v_x}\le v_x$ and the claim holds.
Otherwise, we have seen more than $\nrCounters$ distinct elements, and specifically
\footnotesize
\begin{align}
t > \nrCounters\label{eq:streamLen}.
\end{align}
\normalsize
From Lemma~\ref{lem:sum-of-counters}, it follows that
\ifdefined \NINEPAGES
\small
\fi
\scriptsize
\begin{align}
\min_{y\in C_t} Query(y) \le \frac{t\cdot M \cdotpa{1+\gamma/2}}{\nrCounters}\le\frac{t\cdot M \cdot\eps\cdotpa{1+\gamma/2}}{1+\gamma}.\label{eq:minBound}
\end{align}
\normalsize
Notice that $\forall x\in C_t$,
\ifdefined \EXTENDED
the value of
\fi
\query{} is determined in Line~\ref{line:normalQuery}; that is, $q_{t}(x) = r_{x,t} + s\cdot c_{x,t}$. Next, observe that an item's remainder value is bounded by $\stepLetter-1$ (Line~\ref{line:regularRemainderUpdate} and Line~\ref{line:takeoverRemainderUpdate}). Thus,
\footnotesize
\begin{align}
\forall x,y\in C_t:q_t(x)\ge \stepLetter + q_t(y)\implies c_{x,t} > c_{y,t}.\label{eq:minCondition}
\end{align}
\normalsize
By choosing $y\in\arg\min_{y\in C_t}q_t(y)$, we get that if $v_{x,t}\ge q_t(y) + \stepLetter$, then $q_t(x)\ge q_t(y) + \stepLetter$ and thus $c_{x,t} > c_{y,t}$. Next, we show that if $v_{x,t} \ge t\cdot M \cdot \eps$, then $c_x > \min_{y\in C_t} c_y$ and thus $x$ will never be the ``victim'' in Line~\ref{line:min}:
\scriptsize
\begin{align*}
q_t(x)&\ge v_{x,t}\ge t\cdot M \cdot \eps = t\cdot M \cdot \eps \cdot\frac{1+\gamma/2}{1+\gamma} + M\gamma/2\cdot\frac{t}{\frac{\slack}{\eps}}\\
_\eqref{eq:minBound} &\ge q_t(y) + M\gamma/2\cdot\frac{t}{\frac{\slack}{\eps}}\\
_\eqref{eq:streamLen} &> q_t(y) + M\gamma/2 .
\end{align*}
\normalsize
Next, since $q_t(x)$ and $q_t(y)$ are integers, it follows that
$$q_t(x) \ge q_t(y) + \step = q_t(y)+\stepLetter.$$
Finally, we apply \eqref{eq:minCondition} to conclude that once $x$ arrives with a cumulative volume of $t\cdot M\cdot\eps$, it will never be evicted (Line~\ref{line:min}) and from that moment on its volume will be measured exactly.\qquad
\end{proof}

\subsection*{Proof of Lemma~\ref{lem:runtime}\\}
\begin{proof}
As mentioned before, \rss{} utilizes the SOS data structure that answers queries in $O(1)$.
Updates are a bit more complex as we need to handle weights and thus may be required to move the flow more than once, upon a counter increase.
Whenever we wish to increase the value of a counter (Line~\ref{line:regularCounterUpdate} and Line~\ref{line:takeoverCounterUpdate}), we need to remove the item from its current group and place it in a group that has the increased $c$ value.
This means that for increasing a counter by $n\in\mathbb N$, we have to traverse at most $n$ groups until we find the correct location.
Since the remainder value is at most $\stepLetter-1$ (Line~\ref{line:regularRemainderUpdate} and Line~\ref{line:takeoverRemainderUpdate}), we get that at any time point, a counter is increased by no more than  $\floor{	\frac{\stepLetter - 1 + w}{\stepLetter}}$ (Line~\ref{line:regularCounterUpdate} and Line~\ref{line:takeoverCounterUpdate}).
Finally, since $\stepLetter = \step$, we get that the counter increase is bounded by
	\ifdefined \NINEPAGES
	\small
	\begin{align*}
	\floor{	\frac{\ensuremath{\floor{M\cdot\gamma/2 + 1}} - 1 + w}{\ensuremath{\floor{M\cdot\gamma/2 + 1}}}}< 1 + \frac{w}{M\gamma/2} \le 1 + \frac{2}{\gamma} = O\parentheses{\oneOverG}.\ 
	\normalsize
	\end{align*}
	\fi
	\ifdefined \EXTENDED
    \scriptsize
	\begin{align*}
	\floor{	\frac{\step - 1 + w}{\step}}< 1 + \frac{w}{\frac{M\gamma}{2}} \le 1 + \frac{2}{\gamma} = O\parentheses{\oneOverG}.\quad\qed
	\end{align*}
    \normalsize
	\fi
\end{proof}

\section{Missing Figure}
\label{sec:missing-figure}

Figure~\ref{fig:gamma} shows runtime performance evaluation of \rss{} as a function of $\gamma$ for three different $\varepsilon$ values ($2^{-8},2^{-10},2^{-12}$).
While we indeed obtained a speedup with larger $\gamma$ values, increasing $\gamma$ beyond a certain small threshold has little impact on performance.

\end{document}